\pdfoutput=1 
\documentclass[12pt]{article}

\usepackage{graphicx}
\usepackage{amsmath}
\usepackage{amssymb}
\usepackage{lineno}
\usepackage{url}
\usepackage{xspace,multicol}
\usepackage{siunitx}
\usepackage{subcaption}
\usepackage{color, colortbl}
\usepackage{units}
\usepackage{ragged2e}
\usepackage{array}
\usepackage{tabularx}
\usepackage{authblk}
\usepackage{heppennames}
\usepackage{feynmp}
\usepackage{libertine}
\usepackage{textgreek}
\newcommand{\guineapig}{GuineaPig\xspace}
\newcommand{\sid}{SiD\xspace}
\newcommand{\slic}{\textsc{SLIC}\xspace}
\newcommand{\geant}{\textsc{Geant4}\xspace}
\newcommand{\murm}{%
  \ifmmode
    \mathchoice
        {\hbox{\normalsize\textmu}}
        {\hbox{\normalsize\textmu}}
        {\hbox{\scriptsize\textmu}}
        {\hbox{\tiny\textmu}}%
  \else
    \textmu
  \fi
}
\newcommand{\micron}{\ensuremath{\murm\mathrm{m}\xspace}}

\begin{document}


\title{A Study of the Impact of High Cross Section
ILC Processes on the SiD Detector Design
}

\author{Timothy Barklow\footnote{SLAC National Accelerator Laboratory, 2575 Sand Hill Rd, Menlo Park, CA 94025}, Luc d`Hautuille\footnote{Institute for Particle Physics, 1156 High Street, Santa Cruz, California 95064}, Christopher Milke$^\dagger$,\\Bruce Schumm$^\dagger$, Anne Sch\"utz$^\S$\footnote{Karlsruhe Institute of Technology (KIT), Department of Physics, Institute of Experimental Nuclear Physics (IEKP), Wolfgang-Gaede-Str. 1, 76131 Karlsruhe, Germany}, Marcel Stanitzki\footnote{Deutsches Elektronen-Synchrotron (DESY), Notkestr. 85, 22607 Hamburg, Germany}, Jan Strube\footnote{Pacific Northwest National Laboratory, 902 Battelle Blvd, Richland, WA 99354}}


\maketitle

\begin{abstract}
The SiD concept is one of two proposed detectors to be mounted at the interaction region of the International Linear Collider (ILC). A substantial ILC background arises from low transverse momentum $\Pep\Pem$ pairs created by the interaction of the colliding beams' electromagnetic fields. In order to provide hermeticity and sensitivity to beam targeting parameters, a forward Beamline Calorimeter (BeamCal) is being designed that will provide coverage down to 5 mrad from the outgoing beam trajectory, and intercept the majority of this pair background. Using the SiD simulation framework, the effect of this pair background on the SiD detector components, especially the vertex detector (VXD) and forward electromagnetic calorimeter (FCAL), is explored. In the case of the FCAL, backgrounds from Bhabha and two-photon processes are also considered. The consequence of several variants of the BeamCal geometry and ILC interaction region configuration are considered for both the vertex detector and BeamCal performance.
\end{abstract}


\section{Introduction}
\label{sec:introduction}


Although much more forgiving than for LHC collisions, activity generated by ILC collisions still needs to be considered in the geometric and readout design of SiD detector components, particularly those closest to the beamline. In this study, we have considered the impact of ILC beam-beam collision activity on the pixelization (spatial and temporal) of the pixel vertex detector (VXD), the readout architecture of the electromagnetic calorimeter (ECAL), and the geometry and layout of the Beamline Calorimeter (BeamCal). We have also considered the impact of the longitudinal position of the BeamCal (tied to the value of $L^*$, the distance from the interaction point to the tip of the final focusing quadrupole), and the inclusion of a dedicated magnetic field from an anti-detector-integrated dipole (anti-DiD), intended to sweep beam-beam collision products into the ILC beam exit pipe, on the backgrounds that populate these innermost detector elements.

As currently envisioned, the ILC machine will deliver beam trains at a rate of \unit[5]{Hz}. Three sets of possible beam parameters, as defined in the Technical Design Report~\cite{TDR}, are listed in Table~\ref{tab:ILC_parameters}. At the center-of-momentum energy \mbox{$E_\text{CM} = \unit[500]{GeV}$}, in the energy region for which the ILC would be expected to run in its first years of operation, two sets of parameters are presented. For the baseline (lower luminosity) set, each train would consist of 1312 crossings each separated by \unit[554]{ns}, for a total train duration of 0.72 ms, and with an integrated luminosity of \unit[3.6]{nb$^{-1}$} per train. A higher-luminosity \unit[500]{GeV} scenario envisions a train of 2625 bunches each separated by \unit[366]{ns}, with total train duration of \unit[0.96]{ms} and an integrated luminosity if \unit[7.2]{nb$^{-1}$} per train. A third set of parameters, proposed for running at an upgraded center-of-momentum energy of \unit[1]{TeV}, was not considered in this study. In all cases, though, the interval of \unit[199]{ms} between successive trains can be made use of for digital processing of analog signal pulses. Design considerations due to high cross-section background processes arise both from the effect of individual crossings, as well as from the effect of multiple beam crossings integrated over portions of the, or the entire, beam train.

\begin{table}
\caption{Beam parameters for different phases in the ILC operation scenario (Baseline 500, Luminosity Upgrade, TeV Upgrade)~\cite{TDR}}.
\label{tab:ILC_parameters}
\centering
\begin{tabularx}{\textwidth}{ll|rrr}
\hline\hline
& & \textbf{Baseline 500} & \textbf{Lumi Upgrade} & \textbf{TeV Upgrade}\\
\hline
\cline{1-5}
\hline
E$_{CM}$  &(\si{\GeV}) & 500  & 500  & \num{1000} \\
n$_b$ & & \num{1312} & \num{2625} & \num{2450}  \\
$\Delta t_b$ &(\si{\nano\second}) & 554  & 366   & 366 \\
N & & \num{2.0e10}  & \num{2.0e10}  & \num{1.74e10}  \\
q$_b$ &(\si{\nano\coulomb}) & 3.2  & 3.2  &  2.7  \\
$\sigma_x^*$ &(\si{\nano\metre}) & 474  & 474  &  481 \\
$\sigma_y^*$ &(\si{\nano\metre}) & 5.9 &  5.9  &  2.8 \\
$\sigma_z$ &(\si{\milli\metre}) & 0.3  &  0.3  &  0.25 \\
L &(\si{\per\centi\metre\squared\per\second}) & \num{1.8e34} & \num{3.6e34} & \num{3.6e34} \\
\hline\hline
\end{tabularx}
\end{table}

\section{Simulation of Background Processes}

\subsection{Pair Background}
\begin{figure}
\centering
\begin{subfigure}[t]{0.31\textwidth}
\centering
\includegraphics[width=\textwidth]{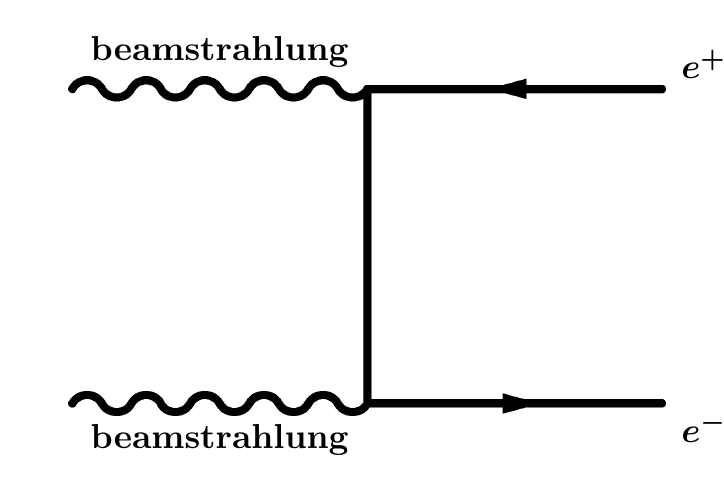}
\caption{Breit-Wheeler}
\end{subfigure}
\hspace*{0.1cm}
\begin{subfigure}[t]{0.31\textwidth}
\centering
\includegraphics[width=\textwidth]{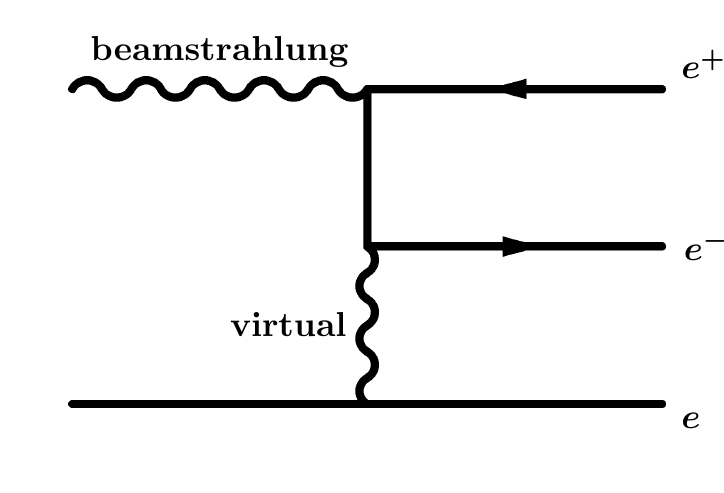}
\caption{Bethe-Heitler}
\end{subfigure}
\hspace*{0.1cm}
\begin{subfigure}[t]{0.31\textwidth}
\centering
\includegraphics[width=\textwidth]{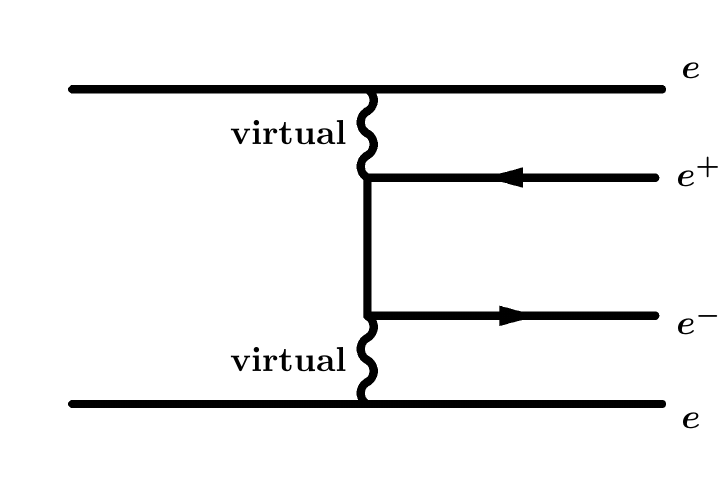}
\caption{Landau-Lifshitz}
\end{subfigure}
\caption{Feynman diagrams of the three LO beamstrahlung pair-production processes.}
\label{fig:beamstrahlung}
\end{figure}

The pair background arises from beam-beam interactions, for which the beamstrahlung from the initial \Pep \Pem beams produces secondary electron-positron pairs. The contributing leading order Feynman diagrams are shown in Figure~\ref{fig:beamstrahlung}.

The pair background events for the results presented in this paper are generated with the Monte Carlo (MC) background event generator \guineapig~\cite{Schulte:1997nga} version 1.4.4. For given accelerator parameters, which have to be provided to \guineapig, the pair background events of one bunch crossing are simulated and stored in an ASCII output file named ``pairs.dat''.
The nominal parameters used for generating the pair background analyzed for this note are given in the Appendix (Section~\ref{sec:Appendix_GuineaPig}). They are entered through the \guineapig input file acc.dat. For reference to the \guineapig manual with the explanations of the parameters see~\cite{GuineaPigMan}.
When running \guineapig, only one bunch worth of background events is generated. By providing different random seeds to \guineapig, it is possible to simulate a full bunch train by letting \guineapig run multiple times. Almost 4000 \guineapig ASCII files of the pair background for the ILC-500 are available on the Grid~\cite{GuineaPigGrid}.
The ASCII output was then converted to one of the following file formats: stdhep or slcio. The conversion tool itself and instructions on its usage are available on the confluence page given in \cite{Confluence}. After the conversion the background events were used as input to a full detector simulation of the \sid detector using the \geant~\cite{geant_ref}~\cite{geant_ref2} based simulation tool \slic~\cite{Graf:2006ei}.


\subsection{Other Background Processes} \label{OBP}

While most of the results in this study rely only on the simulation of incoherent pair backgrounds, one component of the study -- that of the occupancy in the KPiX-instrumented forward electromagnetic calorimeter (FCAL) -- requires the simulation of additional backgrounds, including those from Bhabha scattering, two-photon and electron-photon processes. These processes tend to have much lower cross section and/or activity than those of the pair backgrounds, and are only relevant for the FCAL studies, for which the KPiX chip integrates over an entire bunch train before pushing its data off the detector.

Bhabha events were generated with a virtuality down to $Q^{2} = \unit[1]{GeV^{2}}$, corresponding to an angular cut-off of approximately \unit[4]{mrad}, using the event generator WHIZARD~1.95~\cite{ref:Whizard}. With this cut-off in virtuality, the Bhabha cross section is \unit[278]{nb}, leading to a Bhabha scattering rate of 0.76 events per beam crossing.

 A sample of low $p_\text{T}$, high cross section
 non-perturbative $\Pgg\Pgg \rightarrow$ hadrons events was generated.
 In this sample fluxes of photons from beamstrahlung
 and the peripheral field of individual beam particles (modeled with the
 Weizs\"acker-Williams approximation) were fused into hadronic final states
 using an ad-hoc meson dominance model for $0.3 < m_{\Pgg\Pgg} < \unit[2.0]{GeV}$ ~\cite{Chen:1993dba} and the Pythia MC generator for $m_{\Pgg\Pgg} < \unit[2.0]{GeV}$. These processes contributed approximately 1.2 events per beam crossing.

The WHIZARD~1.95 MC was also used to generate a sample of lower cross section \mbox{$\Pepm \Pgg \rightarrow \Pepm \Pgg$} and \mbox{$\Pepm \Pgg \rightarrow \Pepm \mathrm{f} \overline{\mathrm{f}} $} events (``low cross section events''), where again the initial state photon was either a beamstrahlung or Weizs\"acker-Williams photon. These processes contributed 0.04 events per beam crossing.

\section{The SiD Detector concept}
\label{Detector}

\subsection{SiD in the Detailed Baseline Document}
\begin{figure}
\centering
\includegraphics[angle=90,height=.7\textwidth]{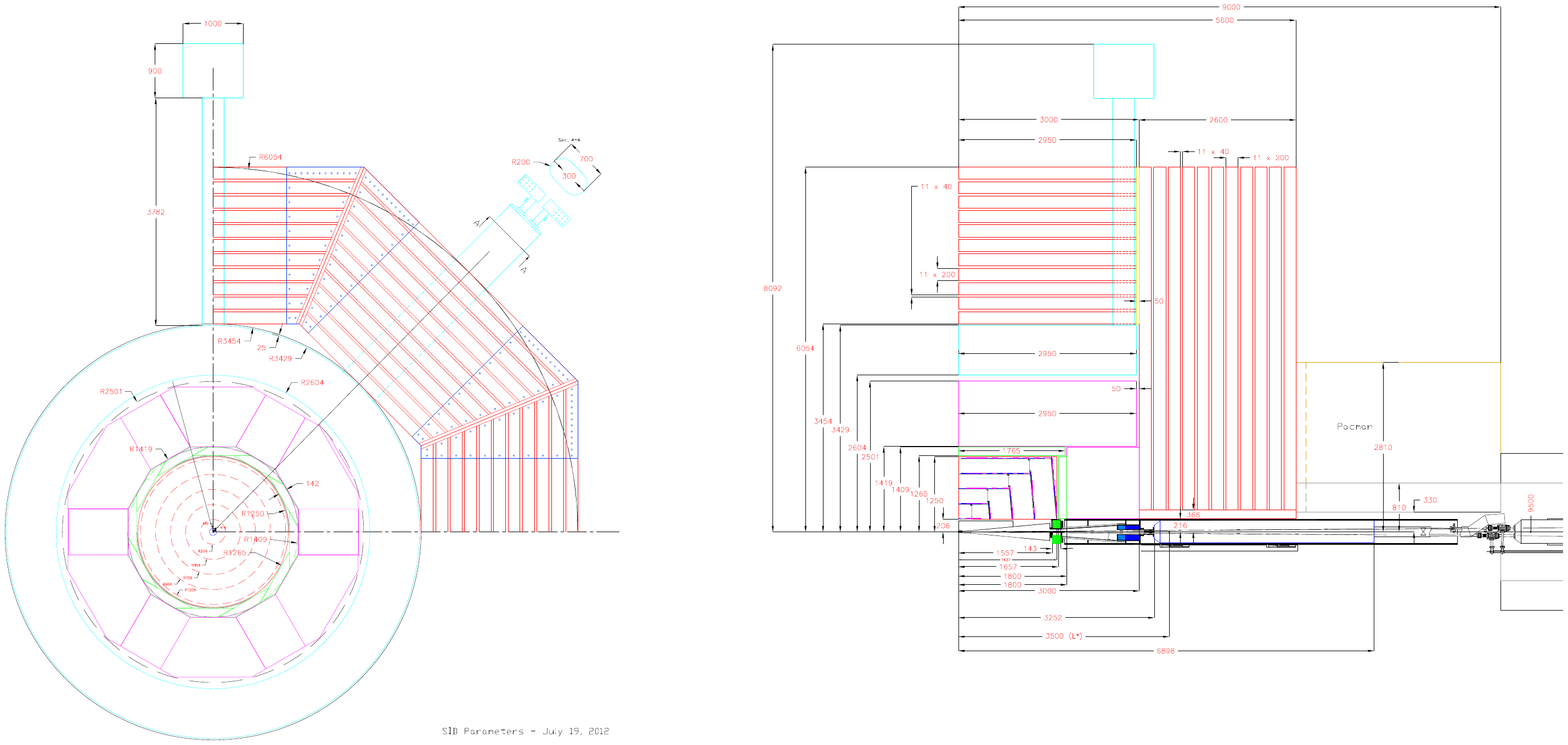}
\caption{Cross sectional views of the SiD concept, as described in the Detailed Baseline Document, in the x-y plane (left), and in the x-z plane (right).}
\label{fig:SiD_Barrel_DBD}
\end{figure}
SiD~\cite{Aihara:2009ad} is a general-purpose detector designed to perform precision measurements at a linear collider. It satisfies the challenging detector requirements for precision measurements at a high-energy electron-positron collider~\cite{ILC_TDR_4}. SiD is optimized for event reconstruction using a particle flow algorithm (PFA), where the reconstruction of both charged and neutral particles is accomplished by the combination of tracking and calorimetry. The net result is a significantly more precise jet energy measurement that results in a di-jet mass resolution good enough to distinguish between W and Z hadronic decays. Following is a description of the SiD detector, as currently envisioned. Some key parameters of the SiD design are presented in Table~\ref{tab:KeyParametersSiD}.

A superconducting solenoid with an inner radius of \unit[2.6]{m} provides a central magnetic field of \unit[5]{T}. The calorimeters are placed inside the coil and consist of a 30 layer tungsten--silicon electromagnetic calorimeter (ECAL) with $\unit[13]{mm^{2}}$ segmentation, followed by a hadronic calorimeter (HCAL) with steel absorber and instrumented with resistive plate chambers (RPC) -- 40 layers in the barrel region and 45 layers in the endcaps. The read-out cell size in the HCAL is $\unit[10\times10]{mm^{2}}$. The iron return yoke outside of the coil is instrumented with 11 RPC layers with $\unit[30\times30]{mm^{2}}$ read-out cells for muon identification.
The silicon-only tracking system consists of five layers of $\unit[20\times20]{\micron^{2}}$ pixels followed by five strip layers with a pitch of \unit[25]{\micron}, a read-out pitch of \unit[50]{\micron} and a length of \unit[92]{mm} per module in the barrel region. The tracking system in the endcap consists of four stereo-strip disks with similar pitch and a stereo angle of $12^\circ$, complemented by four pixelated disks in the vertex region with a pixel size of $\unit[20\times 20]{\micron^{2}}$ and three disks in the far-forward region at lower radii with a pixel size of $\unit[50\times50]{\micron^{2}}$.
Figure~\ref{fig:SiD_Barrel_DBD} shows a schematic drawing of the barrel of SiD, as described in the Detailed Baseline Document.

All sub-detectors have the capability of time-stamping at the level of individual bunches, \unit[366]{ns} apart, $\approx$ 2600 to a train in the luminosity upgrade, according to the ILC TDR. The whole detector will be read out in the \unit[200]{ms} between bunch trains.

\begin{figure}
\centering
\includegraphics[width=.9\textwidth]{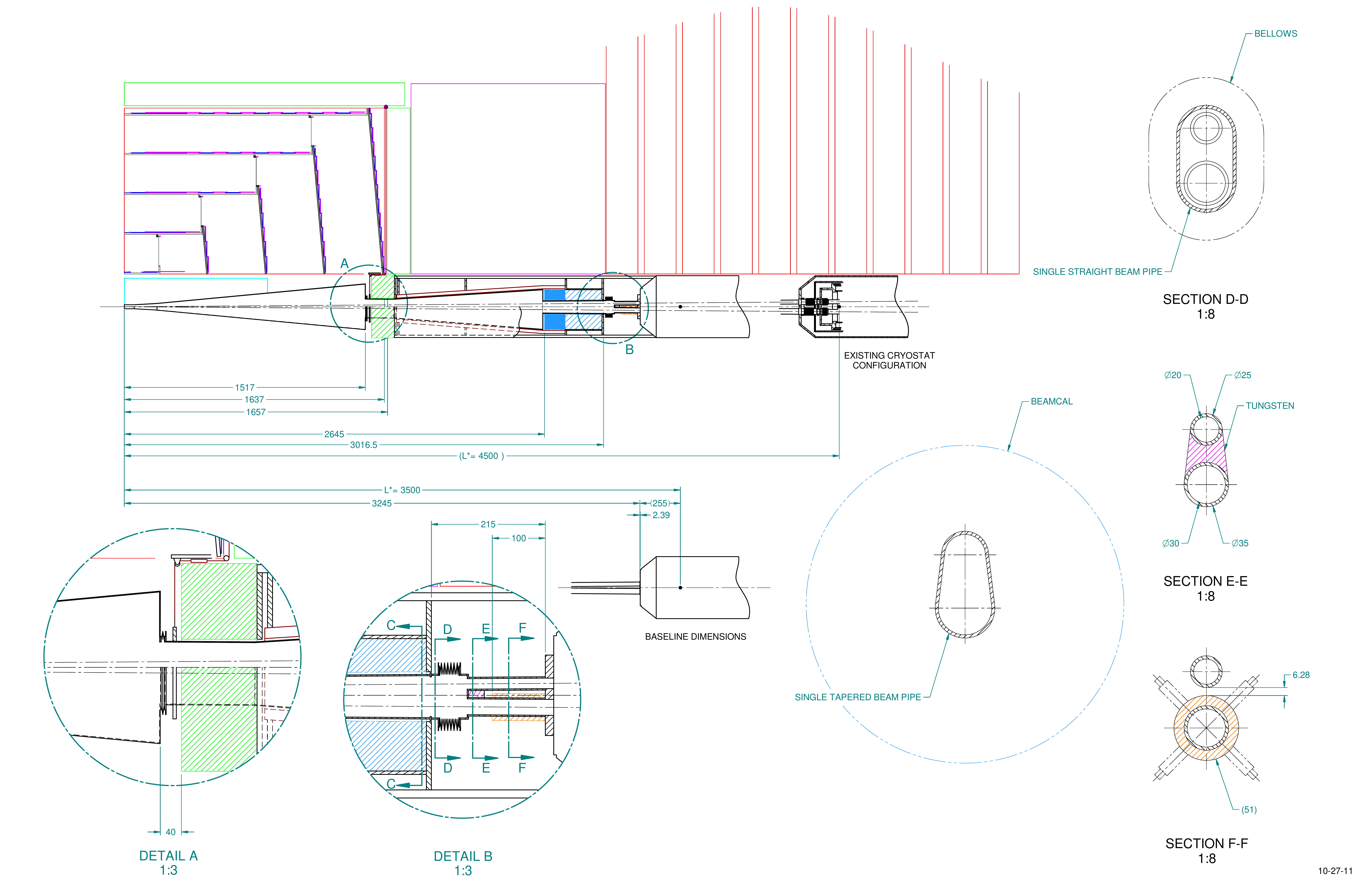}
\caption{Detailed view of the forward layout of the detector with an L* of 4.1 m.}
\label{fig:forward_region_new_Lstar}
\end{figure}
Figure~\ref{fig:forward_region_new_Lstar} shows the detailed layout of the detector in the forward region.
Two dedicated calorimeters are foreseen in the very forward region. The LumiCal, covering an angular range of roughly \unit[40-90]{mrad} from the outgoing beamline, is designed for the precise measurement of the colliding-beam luminosity.  The BeamCal, extending from \unit[40]{mrad} down to angles as small as \unit[5]{mrad}, is designed for fast estimation of the luminosity, and to provide hermetic coverage for high- and intermediate-energy electrons and positrons arising from two-photon processes. The LumiCal and BeamCal are currently envisioned as cylindrical semiconductor-tungsten sampling calorimeters. The BeamCal is placed just in front of the final focus quadrupole and is centered on the outgoing beam. The LumiCal is aligned with the electromagnetic calorimeter endcap. The LumiCal is expected to make use of silicon sensors as the active medium, and is a precision device with challenging requirements on the mechanics and position control. The BeamCal is exposed to a large flux of low-energy electron-positron pairs originating from beamstrahlung. These depositions, useful for a bunch-by-bunch luminosity estimate and the determination of beam parameters, require radiation hard sensors -- as much as \unit[1]{MGy} of electromagnetically induced radiation per year at the peak of the radiation field. The choice of sensor technology is under study with dedicated radiation-damage experiments. The detectors in the very forward region also have to tackle relatively high occupancies, requiring dedicated front-end electronics.

\begin{table}
\caption{Key parameters of the baseline SiD design. (All dimension are given in cm).}
\label{tab:KeyParametersSiD}
\begin{tabular}{lllll}
\hline\hline
SiD Barrel & Technology & Inner radius & Outer radius & z extent \\
\hline
Vertex detector & Silicon pixels & 1.4 & 6.0 & $\pm 6.25$ \\
Tracker & Silicon strips & 21.7 & 122.1 & $\pm 152.2$ \\
ECAL & Silicon pixels-W & 126.5 & 140.9 & $\pm 176.5$ \\
HCAL & RPC-steel & 141.7 & 249.3 & $\pm 301.8$ \\
Solenoid & 5 T SC & 259.1 & 339.2 & $\pm 298.3$ \\
Flux return & Scintillator-steel & 340.2 & 604.2 & $\pm 303.3$ \\
\hline
SiD Endcap & Technology & Inner z & Outer z & Outer radius \\
\hline
Vertex detector & Silicon pixels & 7.3 & 83.4 & 16.6 \\
Tracker & Silicon strips & 77.0 & 164.3 & 125.5 \\
ECAL & Silicon pixel-W & 165.7 & 180.0 & 125.0 \\
HCAL & RPC-steel & 180.5 & 302.8 & 140.2 \\
Flux return & Scintillator/steel & 303.3 & 567.3 & 604.2 \\
LumiCal & Silicon-W & 155.7 & 169.55 &  20.0 \\
BeamCal & Semiconductor-W & 326.5 & 344 & 14.0 \\
\hline\hline
\end{tabular}
\end{table}

\subsection{SiD Detector Variants}

\subsubsection{The Anti-DiD Field}

To potentially suppress BeamCal backgrounds by directing a significant amount of the coherent pair activity into the outgoing beam pipe, the inclusion of an anti-DiD magnetic field has been proposed~\cite{ref:antiDiD}. The resulting vertex detector occupancy and BeamCal reconstruction efficiency, with and without the inclusion of the anti-DiD field, will be discussed below.

\subsubsection{L* Variations}
L* is the distance between the interaction point (IP) and the beginning of the final focusing magnet of the beam line, known as quadrupole QD0. The position of the BeamCal along the z-axis is related to the value of L*, since the BeamCal is attached to the end of the QD0 support structure. Thus, changing L* also changes the BeamCal location. In prior models of the SiD, L* was \unit[3.5]{m}, and the distance between the IP and the BeamCal was 2.95 meters. The most recent (and currently nominal) model of the SiD has L* changed to \unit[4.1]{m}, and has moved the BeamCal to be \unit[3.265]{m} away from the IP, in accordance with the ILC Change Control Board, which has dictated a common L* for the SiD and ILD detectors.

\subsubsection{Variants of the ``plug'' region of the BeamCal}
\begin{figure}[h]
    \centering
    \begin{minipage}{0.3\textwidth}
        \includegraphics[width=\textwidth]{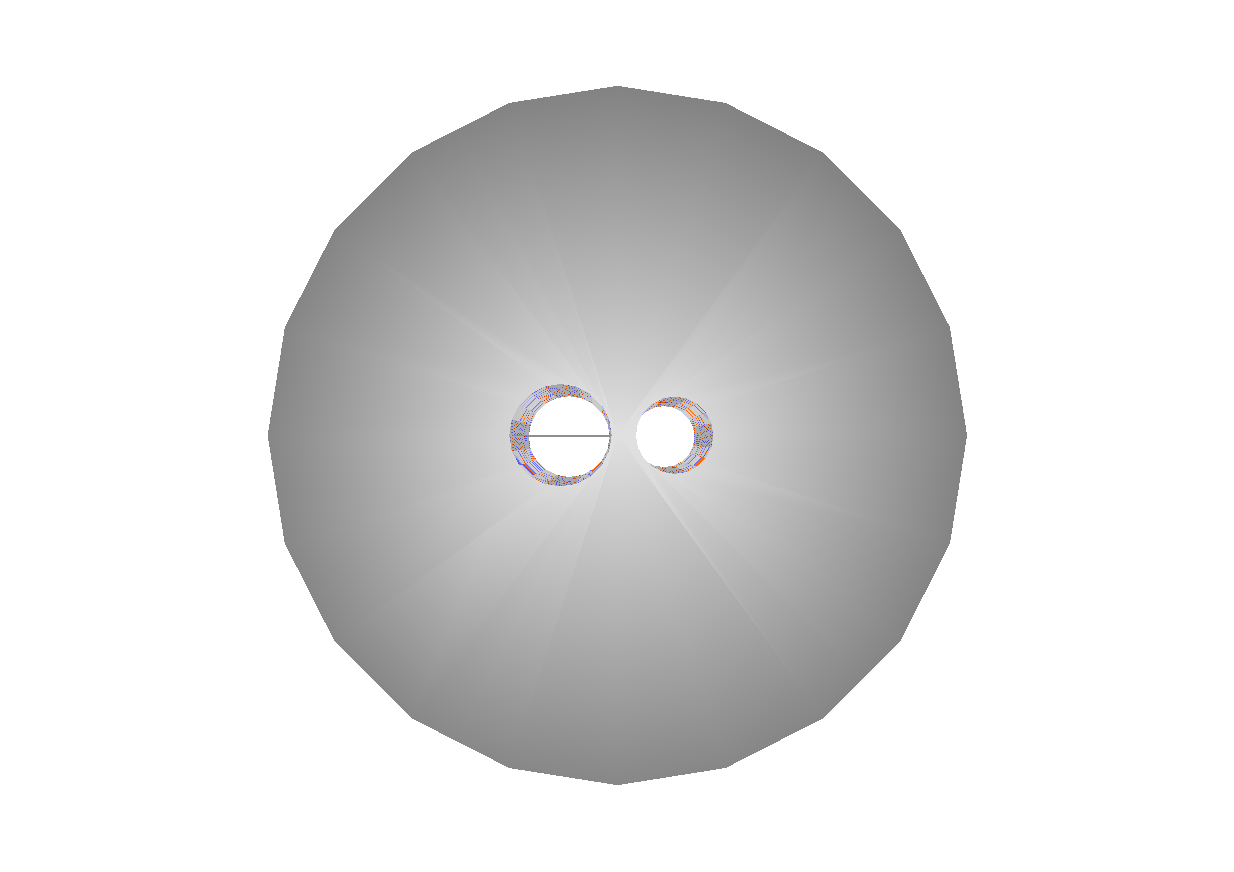}
    \end{minipage}
    \begin{minipage}{0.3\textwidth}
        \includegraphics[width=\textwidth]{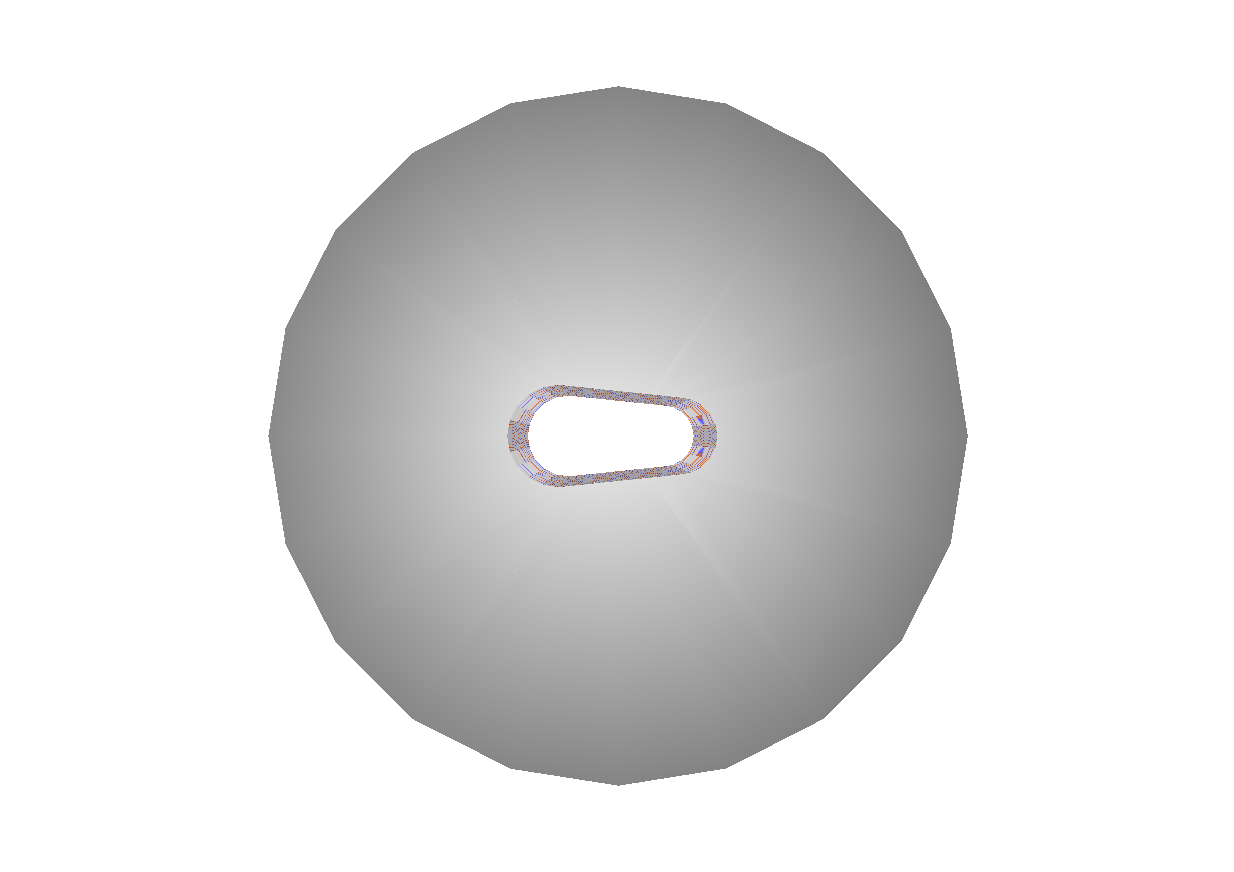}
    \end{minipage}
    \begin{minipage}{0.3\textwidth}
        \includegraphics[width=\textwidth]{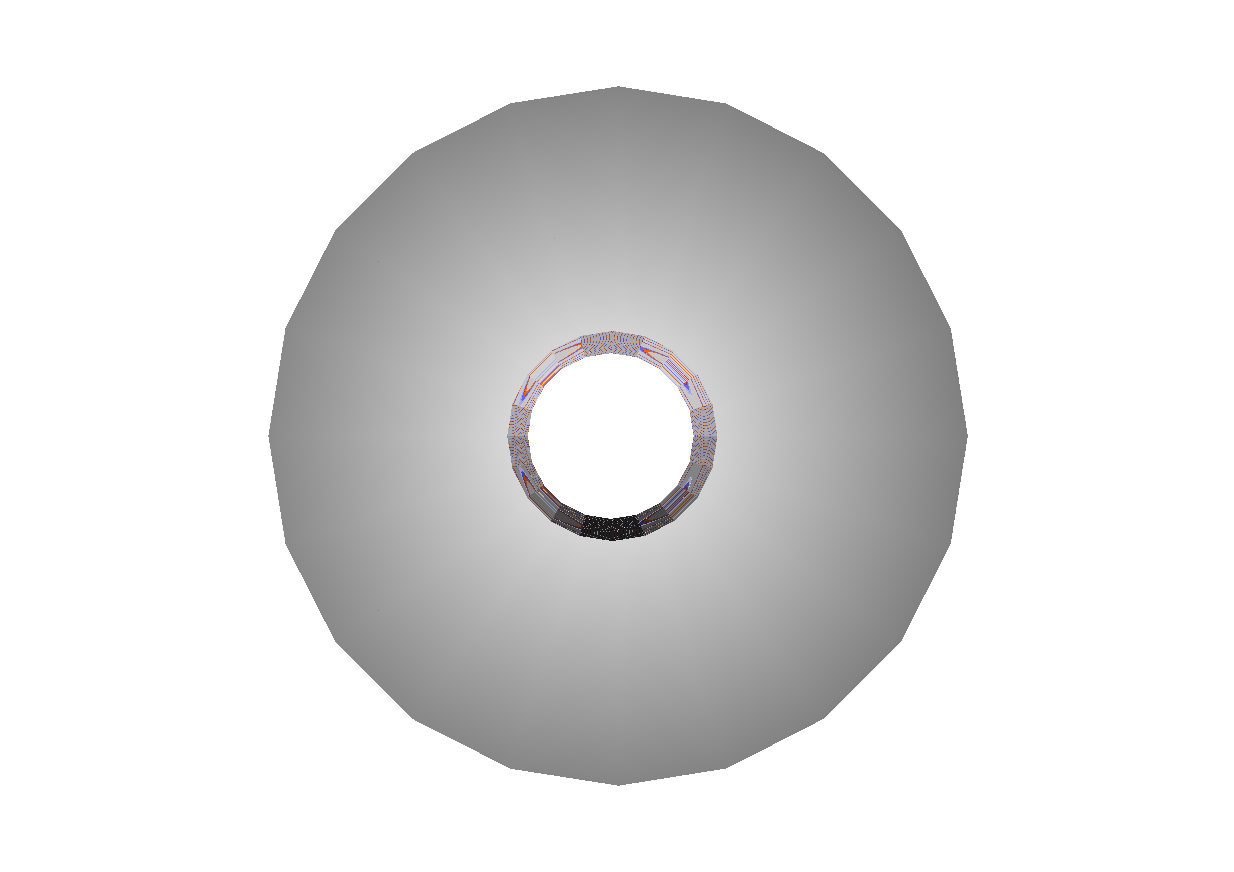}
    \end{minipage}
    \caption{Front face of the BeamCal, showing the three different plug region implementations.}
    \label{fig__beamcal_face}
\end{figure}

The plug region of the BeamCal refers to the area between and immediately surrounding the two beampipe holes. The outgoing beampipe hole, around which the BeamCal is centered, is \unit[20.5]{mm} in radius. The incoming beampipe hole is \unit15.5]{mm} in radius. There are three proposed designs for the plug region, pictured in Figure~\ref{fig__beamcal_face}, which are, from left to right: The first (and currently nominal) design instruments the full plug region, with two holes for the incoming and outgoing beam pipes. The second design is the so-called wedge-cutout, with one cutout for both beam pipes. The final design for the plug region is the circle-cutout, with a diameter which circumscribes both beam pipes. These designs mark a progressively more aggressive approach to removing material from the path of the highest concentration of background energy deposition in the BeamCal. Later sections will discuss the trade-off between BeamCal reconstruction efficiency and the effect of albedo emanating from the BeamCal associated with these variants.

\section{Channel Hit Counts in the Forward Electromagnetic Calorimeter}

In its baseline design, the segments of the SiD calorimeter (described briefly in Section~\ref{Detector}), will be read out with the kPiX chip~\cite{KPiX}. In the current kPiX design, the kPiX chip accumulates hit information over the course of a full bunch train, storing the charge from each hit on a buffer capacitor. After the bunch train has ended, the stored charge is digitized and the result encoded and transmitted into the data acquisition system. Thus, if there are more hits in a given channel during the train than there are buffer capacitors for that channel (in the current design the buffer depth is four), information is lost. The following study addresses this concern, simulating the accumulated effects of all four background types (pair, Bhabha, two-photon, and high cross-section events). The study has been done for the high-luminosity \mbox{$\mathrm{E}_\text{CM} = \unit[500]{GeV}$} scenario, which is specified in the second column of Table~\ref{tab:ILC_parameters}.

For each of the pair, Bhabha, two-photon and low-cross-section backgrounds (see Section~\ref{OBP}), the number of events corresponding to the integrated luminosity of one full train of beam crossings is analyzed, and hits are accumulated in the channels of the FCAL sensors.
For each such process, the number of included events was chosen according to a Poisson probability distribution with a mean given by the expected number of events per train. The hits from these events were then combined with those from the other three background processes.

\begin{figure}
\centering
\includegraphics[width=0.8\textwidth]{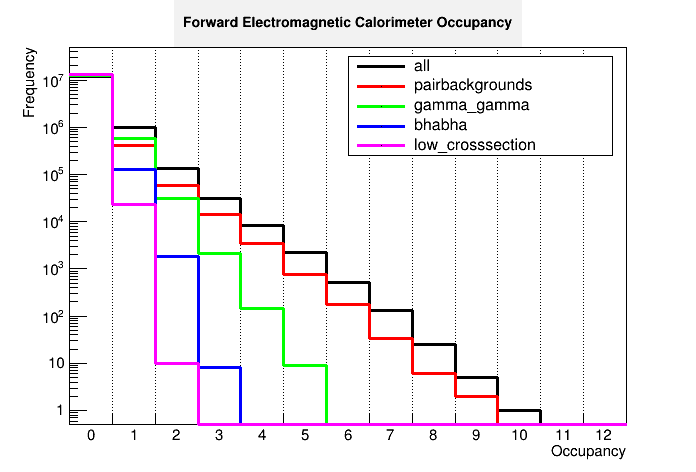}
\caption{Distribution of number of hits collected in each channel of the FCal during a full train of beam crossings. The individual contributions of the four different background sources are shown in different colors; their combined effect is shown in black.}
\label{fig:channel_count}
\end{figure}

\begin{figure}
\centering
\includegraphics[width=0.8\textwidth]{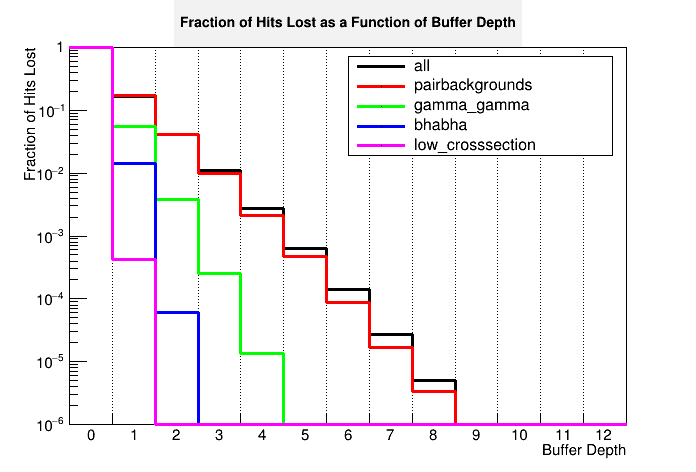}
\caption{Dependence of the overall fraction of FCAL hits lost as a function of kPiX buffer depth. The individual contributions of the four different background sources are shown in different colors; their combined effect is shown in black.}
\label{fig:frac_loss}
\end{figure}

Figure~\ref{fig:channel_count} shows the distribution of number of hits per train, for each individual background source and for all sources together, over the entire FCAL. As many as ten hits fall on any given channel during a typical train. Greater than one in ten thousand channels accumulates five or more hits during the train, at which point the current design with four buffers begins to suffer a loss of information. Correspondingly, Figure~\ref{fig:frac_loss} shows the fraction of hits lost as a function of buffer depth. The simulation suggests that with a buffer depth of four, a few tenths of a percent of FCAL hits will be lost in any given train of beam crossings. With a buffer depth of six this fraction drops to just above $10^{-4}$ and with a buffer depth of eight the fraction of hits lost drops to below $10^{-5}$.

\begin{figure}
\centering
\includegraphics[width=0.8\textwidth]{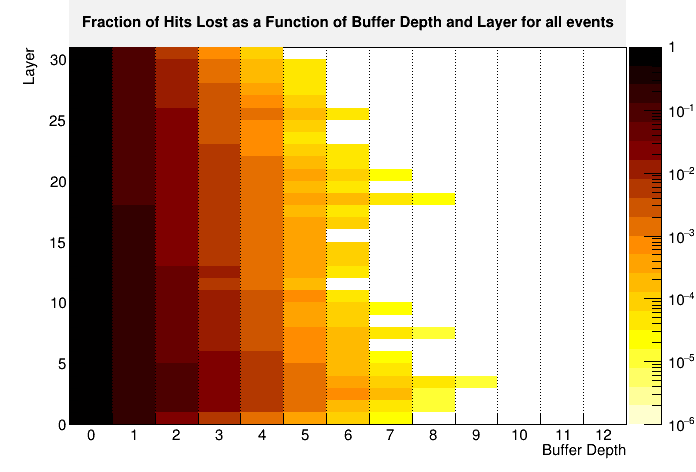}
\caption{Dependence of the fractional loss of hits on buffer depth and depth within the FCAL (layer).}
\label{fig:frac_loss_layer}
\end{figure}

\begin{figure}
\centering
\includegraphics[width=0.8\textwidth]{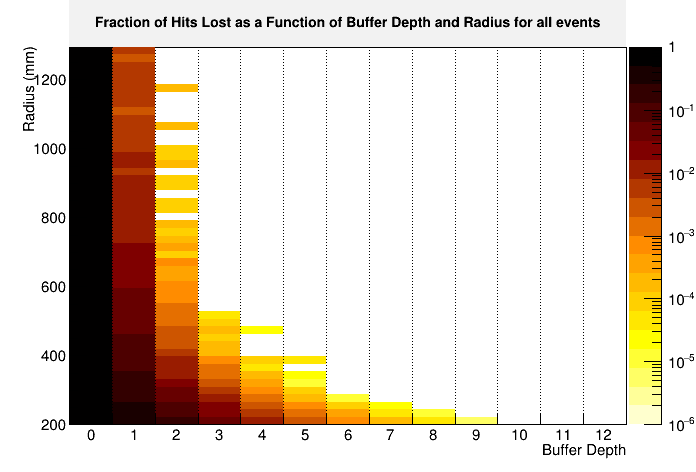}
\caption{Dependence of the fractional loss of hits on buffer depth and transverse distance from the beam line.}
\label{fig:frac_loss_radius}
\end{figure}

However, this loss rate may have significant dependencies on the depth within the calorimeter (due to the great density of hits at shower maximum) and on the distance from the beam line. Figure~\ref{fig:frac_loss_layer} shows the fraction of hits lost as a function of buffer depth and layer number (depth within the calorimeter). For a buffer depth of four, the mean fractional hit loss at shower max for a high-energy electromagnetic particle (roughly layer 10) is a few tenths of a percent, and drops to below $10^{-3}$ for a buffer depth of six.
Figure~\ref{fig:frac_loss_radius} shows the dependence of the fractional loss of hits on buffer depth and transverse distance from the beam line. Because the backgrounds are dominated by $t$-channel processes, a significant radial dependence is observed. For a buffer depth of four, about one percent of hits are lost at the innermost radius of 200 mm, which is significantly above the average hit loss. Using six buffers in the kPiX, however, reduces the fractional hit loss to below $10^{-3}$ at any radius.

\section{Vertex Detector}
\subsection{Vertex Detector Occupancies}
\label{VXD_Occ}


The occupancy of a given detector element, or sub-region within a detector element, is defined as the fraction of channels (pixels in the case of the ILC vertex detector (VXD)) that are hit over a given accumulation of beam crossings relative to the total number of channels in that detector element or region. While no strict requirements present an upper limit on the tolerable level of occupancy, values above $10^{-2}$ are generally considered unacceptable for tracking systems, while values below $10^{-3}$ are considered acceptable for any detector system.

Precision tracking -- both in terms of pattern recognition as well as the accurate measurement of the trajectories of the individual particles arising from beam collisions -- is the most susceptible to degradation from high background occupancy. Since the most sensitive and precise SiD tracking subsystem -- the vertex detector -- is also the closest tracking system to the ILC beam line, where beam-induced backgrounds are expected to be largest, a study has been undertaken of the VXD occupancy and its dependence upon various ILC and SiD design choices, including the value of L*, the geometry of the BeamCal, and the spatial and temporal granularity of the VXD sensors.

\begin{figure}
\centering
\includegraphics[width=0.8\textwidth]{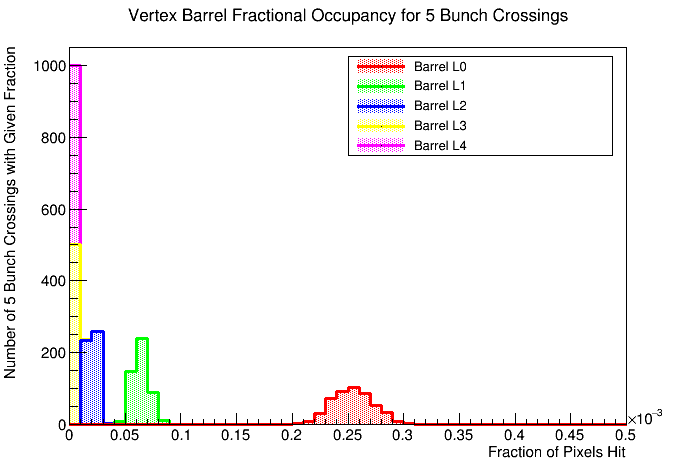}
\caption[Beamstrahlung processes]{VXD barrel occupancies by layer, assuming a pixel size of $\unit[30 \times 30]{\micron^{2}}$ and an integration time of five bunch crossings.}
\label{fig:barrel_occ_all}
\end{figure}

\begin{figure}
\centering
\includegraphics[width=0.8\textwidth]{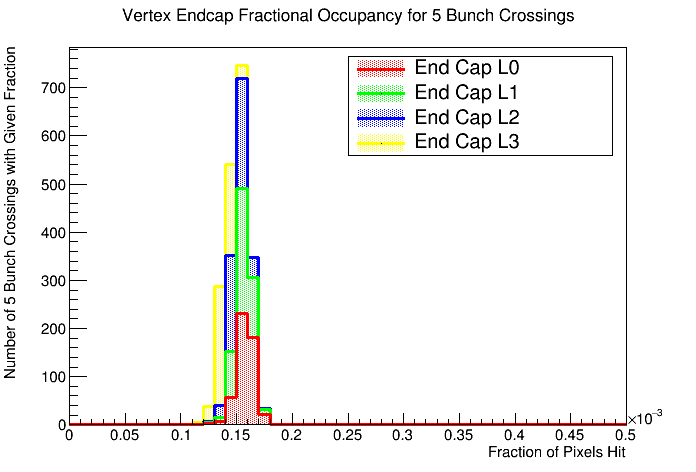}
\caption[Beamstrahlung processes]{VXD endcap occupancies by layer, assuming a pixel size of $\unit[30 \times 30]{\micron^{2}}$ and an integration time of five bunch crossings.}
\label{fig:endcap_occ_all}
\end{figure}

Figures~\ref{fig:barrel_occ_all} and~\ref{fig:endcap_occ_all} show the occupancy distribution for the nominal SiD/L* configuration and for the most conservative granularity considered for the VXD: a pixel size of $\unit[30 \times 30]{\micron^{2}}$ and an integration time of 5 bunch crossings. Occupancies are shown averaged over each layer of the barrel and endcap regions of the VXD. Layer-averaged occupancies are well below $10^{-3}$ in all cases.

\begin{figure}
\centering
\includegraphics[width=0.8\textwidth]{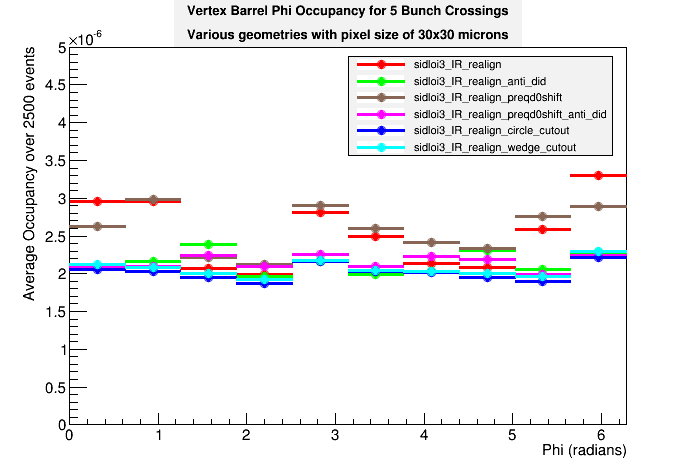}
\caption[Beamstrahlung processes]{VXD barrel layer 0 occupancies as a function of azimuthal angle, assuming a pixel size of $\unit[30 \times 30]{\micron^{2}}$ and an integration time of five bunch crossings.}
\label{fig:barrel_0}
\end{figure}

\begin{figure}
\centering
\includegraphics[width=0.8\textwidth]{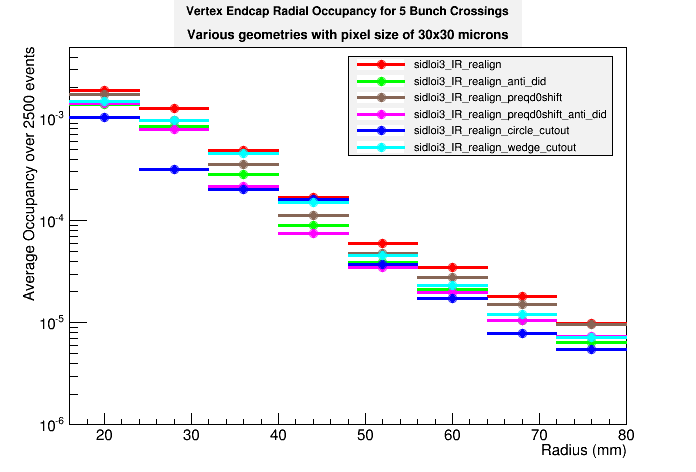}
\caption[Beamstrahlung processes]{VXD endcap occupancies as a function of radius from the beamline, assuming a pixel size of $\unit[30 \times 30]{\micron^{2}}$ and an integration time of five bunch crossings.}
\label{fig:endcap_0}
\end{figure}

However, some caution must be taken, since it is possible that the occupancy can vary locally within any given layer. To this end, the layer with the highest mean occupancy (layer 0 for both the barrel and endcap) is examined further to determine whether there are regions that have potentially unacceptably high occupancy. Figure~\ref{fig:barrel_0} shows the barrel layer 0 occupancy as a function of azimuthal angle (averaging over the $z$ extent of the barrel); the occupancy is seen to remain well below $10^{-3}$ for any value of $\phi$. Figure~\ref{fig:endcap_0} shows the endcap layer 0 occupancy as a function of radius from the beamline (averaging over the azimuthal angle). At all but the innermost radii, the occupancy remains below $10^{-3}$, and never rises above $2\times10^{-3}$. 

In addition to the azimuthal and radial occupancy dependencies for the nominal SiD/L* configuration, studies have also been made for alternative configurations. Since the predominant source of VXD backgrounds is expected to arise from the albedo emanating from the front face of the BeamCal, the occupancy has also been simulated for BeamCal geometries for which the wedge between the outgoing and incoming stay-clear holes is removed, as well as for the case for which the circle cutout is removed (see Section~\ref{Detector} for a description of the wedge and circle cutouts). Studies are also shown for the cases of a reduced L* of \unit[3.5]{m}, and for the inclusion of the anti-DiD field. All in all, occupancies show little dependence on the choice of configuration, and remain comfortably small even for the conservative choice of VXD pixelation.

\subsection{Time distribution of hits in the vertex detector}
\label{VXD_time}
Background particles are created at the time of the bunch crossing. However, the background particles do not all hit the subdetectors at the same time shortly after the bunch crossing. This is due to the fact that some of the particles spiral in the solenoid field or scatter off other detector components, and therefore arrive at the detector component significantly later than the instant of the bunch crossing. This can be taken advantage of to reduce detector background levels by applying electronic/digital gates that cut away the late arriving background particles.

As will be illustrated below, the observed features of the hit-time distributions lead to the conclusion that some of the late particles can be clearly separated in time from the original background particles.
The following distributions make use of a time variable measured relative to the instant of the bunch crossing, and will show that some of the background particles will be several 100 nanoseconds to several microseconds late. It should be noted that for the ILC \unit[500]{GeV}\cite[Volume 1, p. 11]{TDR}, the temporal spacing between successive bunch crossings is \unit[554]{ns}.

\begin{figure}
\centering
\includegraphics[width=0.8\textwidth]{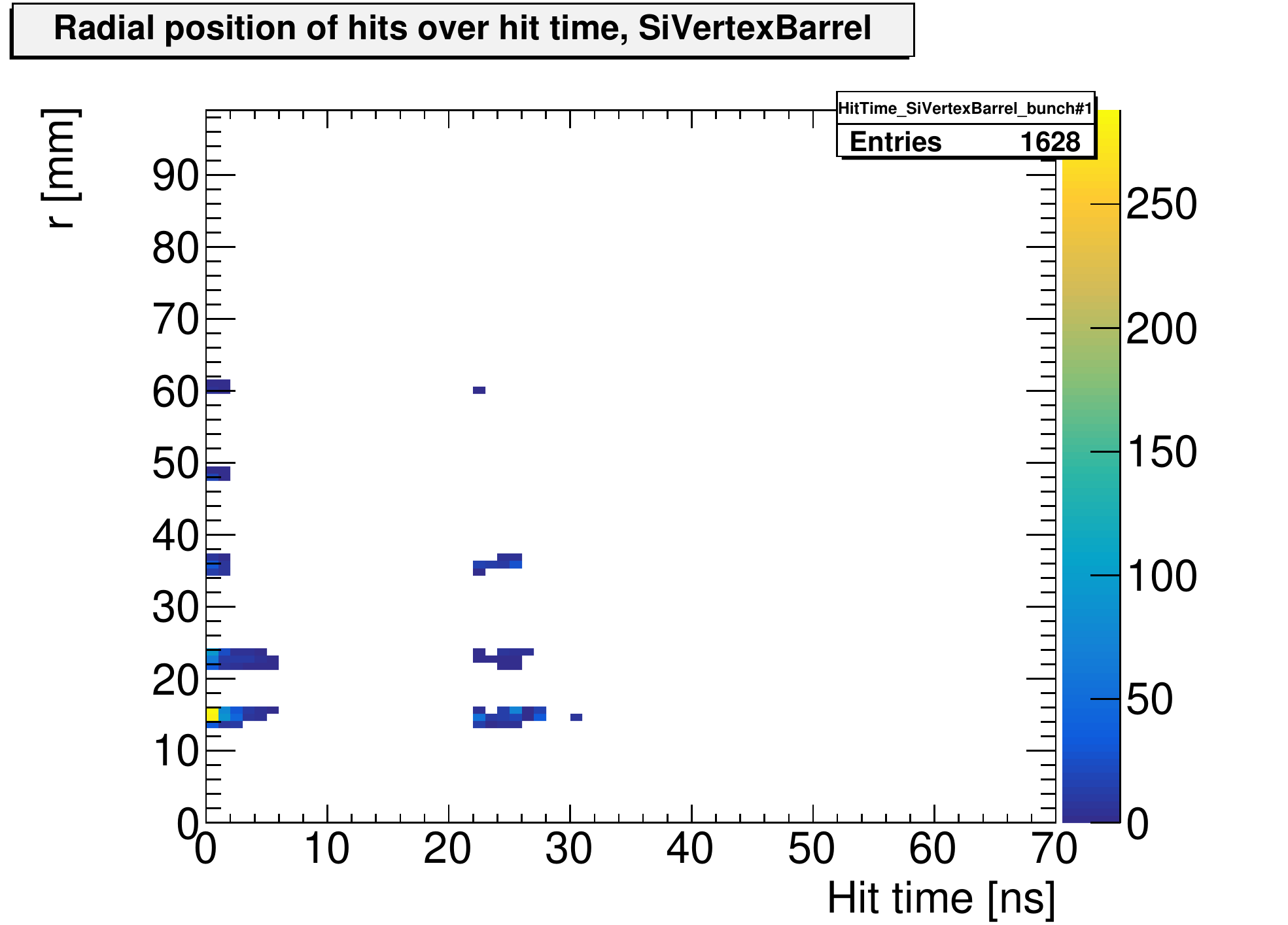}
\caption[Hit time distribution of one bunch in the barrel of the SiD vertex detector]{Distribution of time of hits induced by pair background particles from one bunch crossing in the barrel of the vertex detector. The plot shows the radial position of hits as a function of the time relative to the bunch crossing.}
\label{fig:hittime_VtxBarrel_1bunch}
\end{figure}
\begin{figure}
\centering
\includegraphics[width=0.8\textwidth]{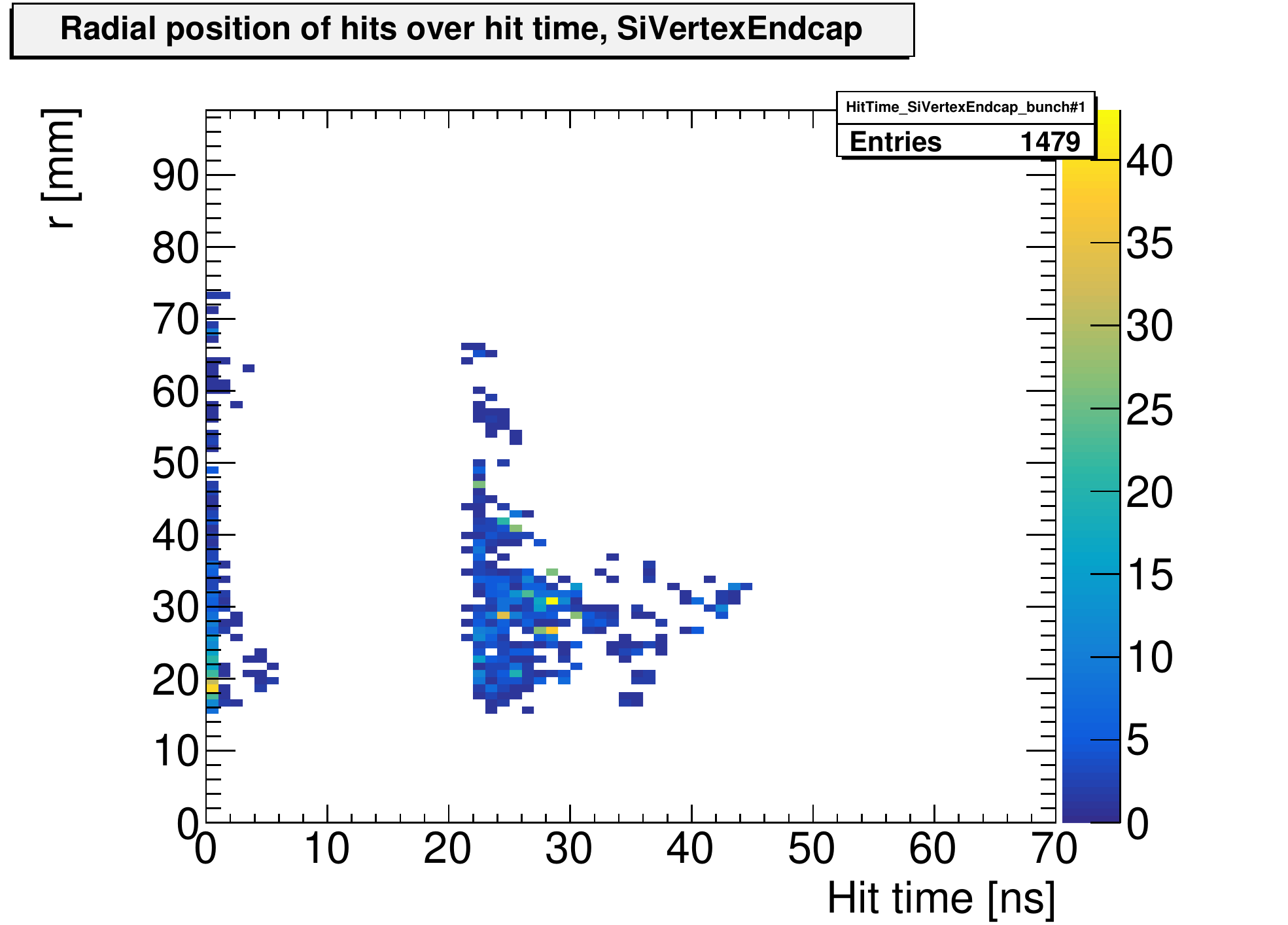}
\caption[Hit time distribution of one pair background bunch in the endcaps of the SiD vertex detector]{Distribution of the time of hits induced by pair background particles from one bunch in the endcaps of the vertex detector. The plot shows the radial position of hits as a function of the time relative to the bunch crossing.}
\label{fig:hittime_VtxEndcap_1bunch}
\end{figure}
\begin{figure}
\centering
\includegraphics[width=0.8\textwidth]{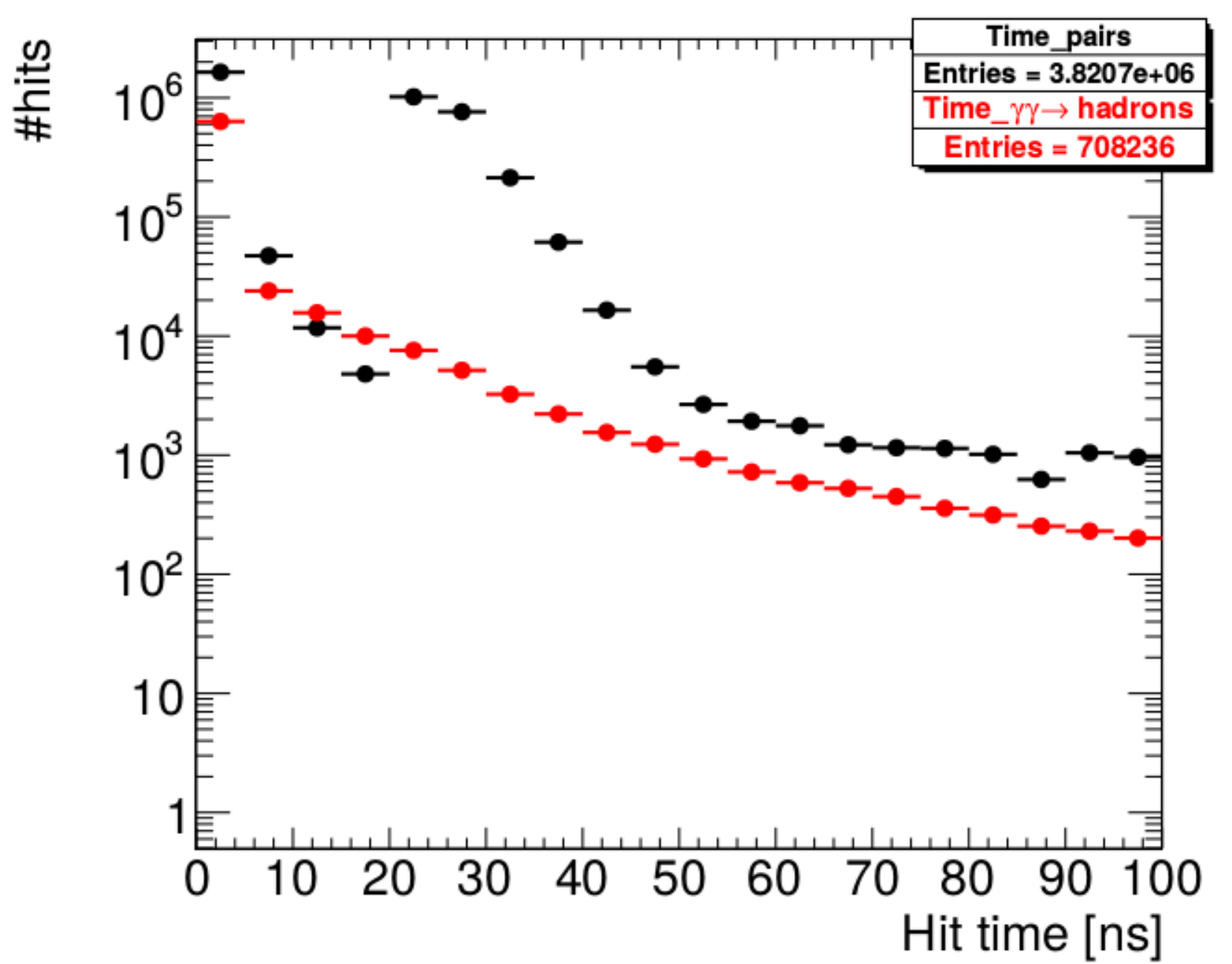}
\caption[Comparison of the hit time distributions of the $\Pgg\Pgg \rightarrow$ hadrons events and the pair background from one train in the barrel and endcaps of the SiD vertex detector]{Distributions of the time of hits in the barrel and endcaps of the vertex detector induced by pair background and the $\Pgg\Pgg \rightarrow$ hadrons events from one train. The hits from the $\Pgg\Pgg \rightarrow$ hadrons events are instantaneous and decrease in number continuously over time. In contrast to that, the hit time distribution of the pair background shows a dip between 10 and \unit[20]{ns}, which was discovered in the previous figures and which will be explained in this Section~\ref{VXD_time}.}
\label{fig:gg-hadrons_pairs_hittime_VtxEndcapBarrel}
\end{figure}

Figures~\ref{fig:hittime_VtxBarrel_1bunch} and \ref{fig:hittime_VtxEndcap_1bunch} show the radial position of the pair background hits as a function of the hit time in the barrel and the endcaps of the vertex detector, respectively, for a single bunch crossing.
In the plot for the vertex barrel (Figure~\ref{fig:hittime_VtxBarrel_1bunch}), the different barrel layers are clearly visible. The hits in the endcaps (Figure~\ref{fig:hittime_VtxEndcap_1bunch}), on the other hand, are continuously distributed in radius. For both distributions there is a gap in time, during which very few particles hit the subdetectors.
Figure~\ref{fig:gg-hadrons_pairs_hittime_VtxEndcapBarrel} now shows a histogram of the hit time distribution for the pair background in the vertex barrel and endcaps for 1312 bunch crossings, i.e. for a full train. The dip in the number of hits between 10 and \unit[20]{ns}, that was already visible in Figures~\ref{fig:hittime_VtxBarrel_1bunch} and \ref{fig:hittime_VtxEndcap_1bunch} before, is very distinct in this plot, and will be explained later. The comparison to the time distribution induced by the high cross-section background, i.e. the $\Pgg\Pgg \rightarrow$ hadrons events, suggests that the vast majority of these hits are instantaneous, and the number of hits drops continuously as expected. As a result of this, the time distribution of the pair background will be the only focus of this section.\\
To explain the anomalous time distribution of the pair background, the following histogram will look at the origins of the background particles, their time of creation, and their momenta.

\begin{figure}
\begin{subfigure}[t]{0.5\textwidth}
\centering
\includegraphics[width=\textwidth]{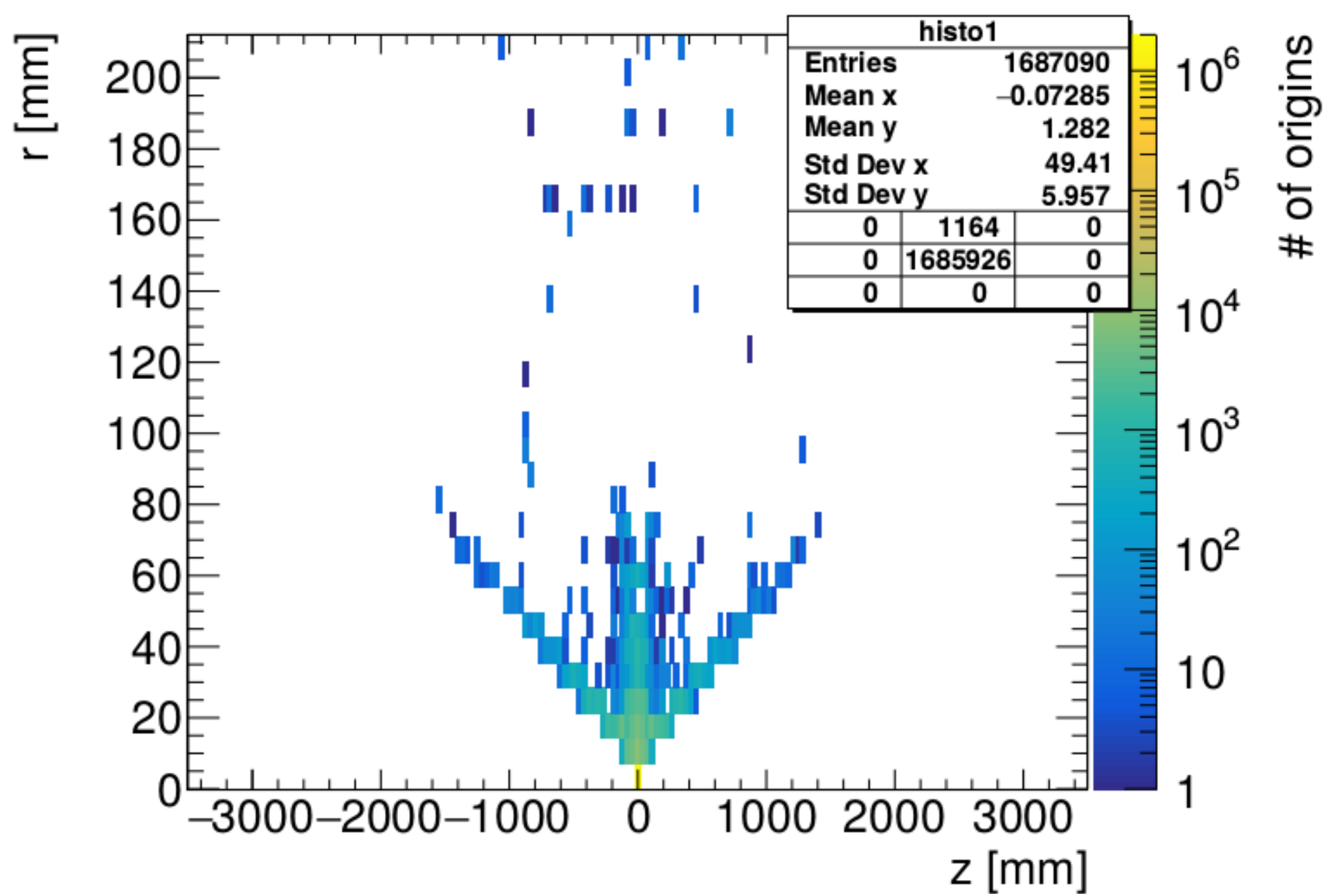}
\caption{Hit time: [\unit[0]{ns};\unit[10]{ns}]}
\end{subfigure}
\begin{subfigure}[t]{0.5\textwidth}
\centering
\includegraphics[width=\textwidth]{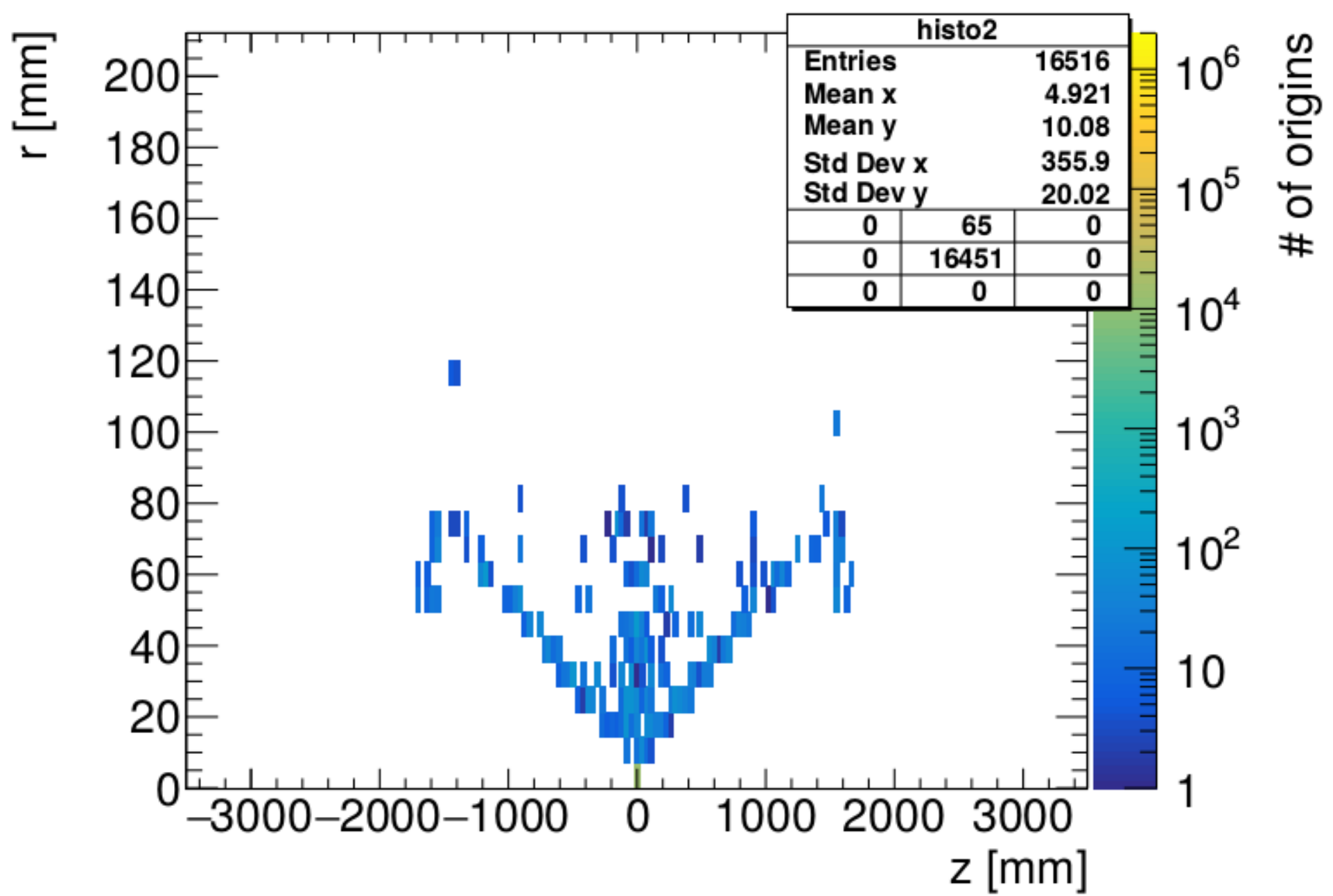}
\caption{Hit time: ]\unit[10]{ns};\unit[20]{ns}]}
\end{subfigure}
\begin{subfigure}[t]{0.5\textwidth}
\centering
\includegraphics[width=\textwidth]{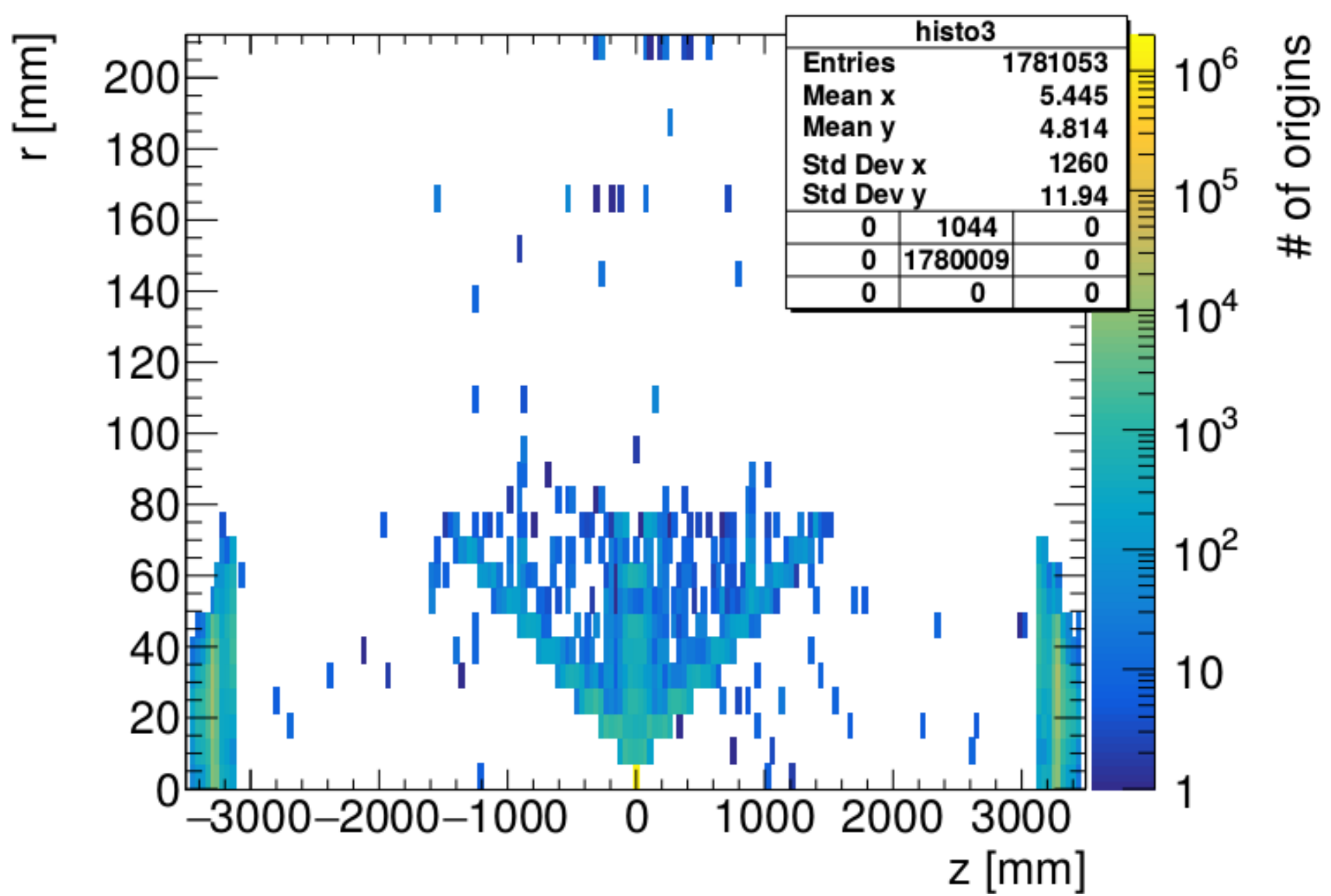}
\caption{Hit time: ]\unit[20]{ns};\unit[30]{ns}]}
\end{subfigure}
\begin{subfigure}[t]{0.5\textwidth}
\centering
\includegraphics[width=\textwidth]{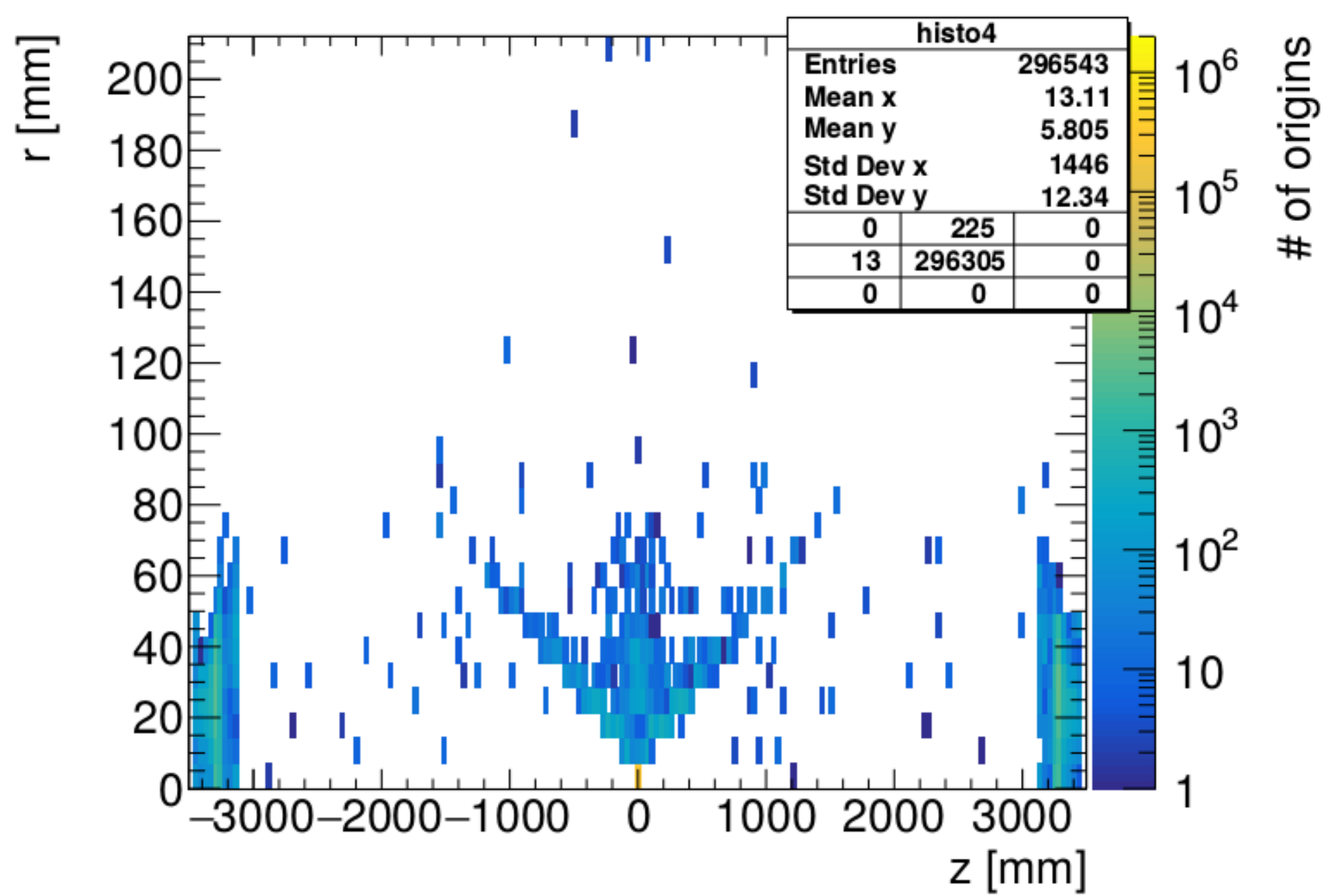}
\caption{Hit time: ]\unit[30]{ns};\unit[50]{ns}]}
\end{subfigure}
\caption[Origins of pair background particles hitting the barrel and the endcaps of the SiD vertex detector]{Point of origin of pair background particles that hit the barrel and endcaps of the vertex detector, arriving at the vertex detector in the time intervals [\unit[0]{ns};\unit[10]{ns}], ]\unit[10]{ns};\unit[20]{ns}], ]\unit[20]{ns};\unit[30]{ns}] and ]\unit[30]{ns};\unit[50]{ns}].
To increase statistics, the plots are averaged over 1312 bunches.}
\label{fig:particleorigin_vertexendcap}
\end{figure}

The subfigures in Figure~\ref{fig:particleorigin_vertexendcap} show the origin of the pair background particles hitting the vertex barrel and the endcaps, divided into the hit time intervals $[\unit[0]{ns};\unit[10]{ns}]$, $]\unit[10]{ns};\unit[20]{ns}]$, $]\unit[20]{ns};\unit[30]{ns}]$ and $]\unit[30]{ns};\unit[50]{ns}]$. The time shown is the time at which the particles are hitting the vertex detector. To increase the statistics, the distributions were made for a full train of 1312 bunches. As the event record stores information about the point of origin of particles arising within a radius that only includes the vertex, tracker, and forward components, the plots are blank in the regions of the calorimeters. Statements about the rough origin of backscatter particles can be made nonetheless.

In Figures~\ref{fig:hittime_VtxBarrel_1bunch} and \ref{fig:hittime_VtxEndcap_1bunch}, the first set of background particles hitting the vertex detector, arriving between 0 and \unit[10]{ns}, is clearly separated in time from the next set, which arrives between 20 and \unit[50]{ns}. The time-sliced plots of the particle origins in Figure~\ref{fig:particleorigin_vertexendcap} show that initially the particles come only from the center of the detector, but later on they also originate in the endcaps of the SiD detector. The particles traveled from the interaction region towards the SiD detector endcaps and the BeamCal calorimeter, and backscatter there. These backscatter particles together with particles newly created in the endcaps and BeamCal subdetector make their way back towards the center of the detector, and hence towards the vertex detector.

\begin{figure}
\centering
\includegraphics[width=0.85\textwidth]{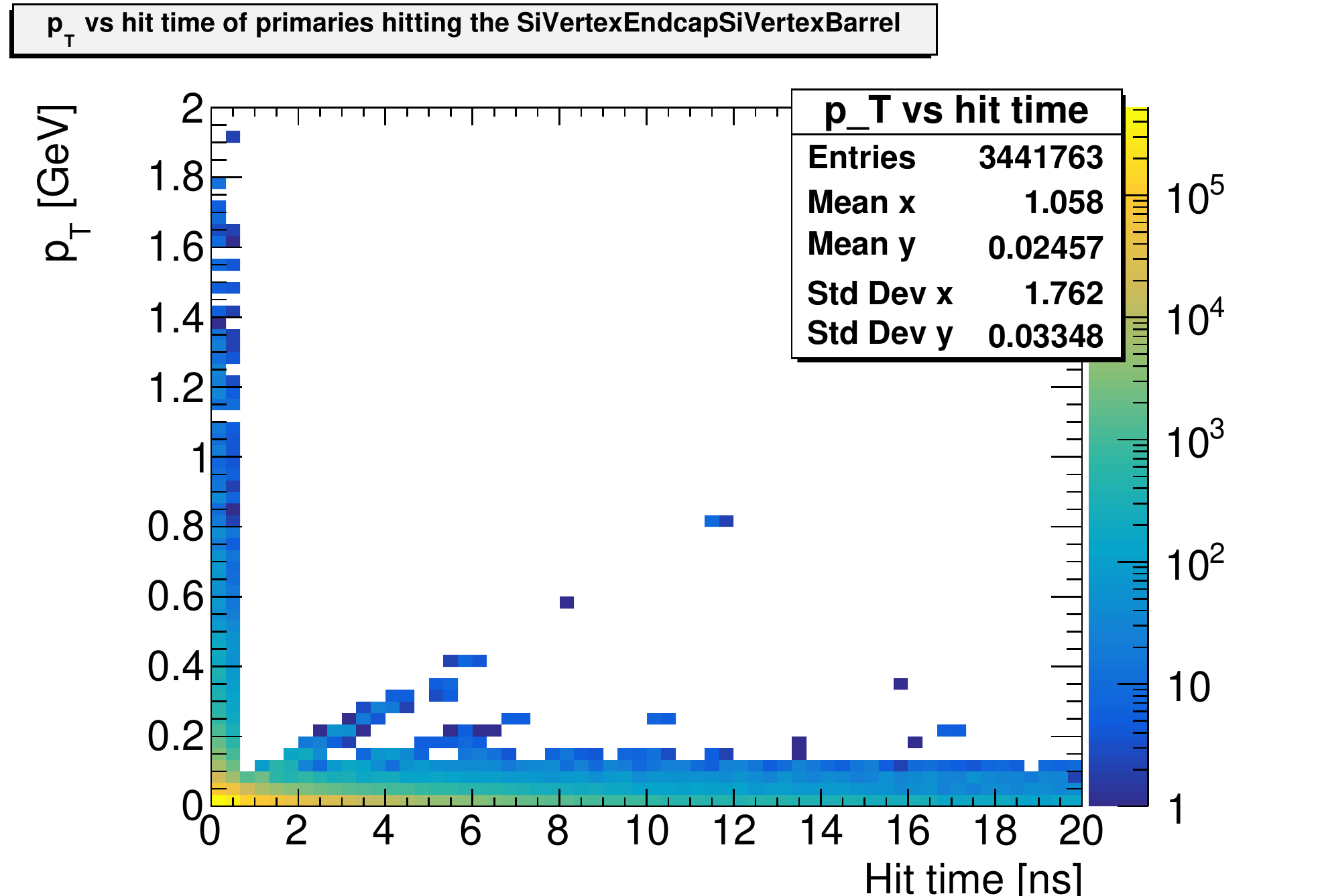}
\caption[$p_T$ vs hit time of primary pair background particles hitting the SiD vertex detector]{The transverse momentum of primary pair background particles created at the IP as a function of the time of the hit in the vertex endcaps and barrel. $\sim$68\% of all primaries are hitting the vertex detector later than \unit[1]{ns} after the bunch crossing, and have a $p_T$ lower than \unit[0.2]{GeV}. Only for very few primary particles the transverse momentum reaches a value of up to $\sim$\unit[0.4]{GeV}.}
\label{fig:PT_hittime_VtxEndcapBarrel}
\end{figure}

\begin{figure}
\begin{subfigure}[t]{0.5\textwidth}
\centering
\includegraphics[width=\textwidth]{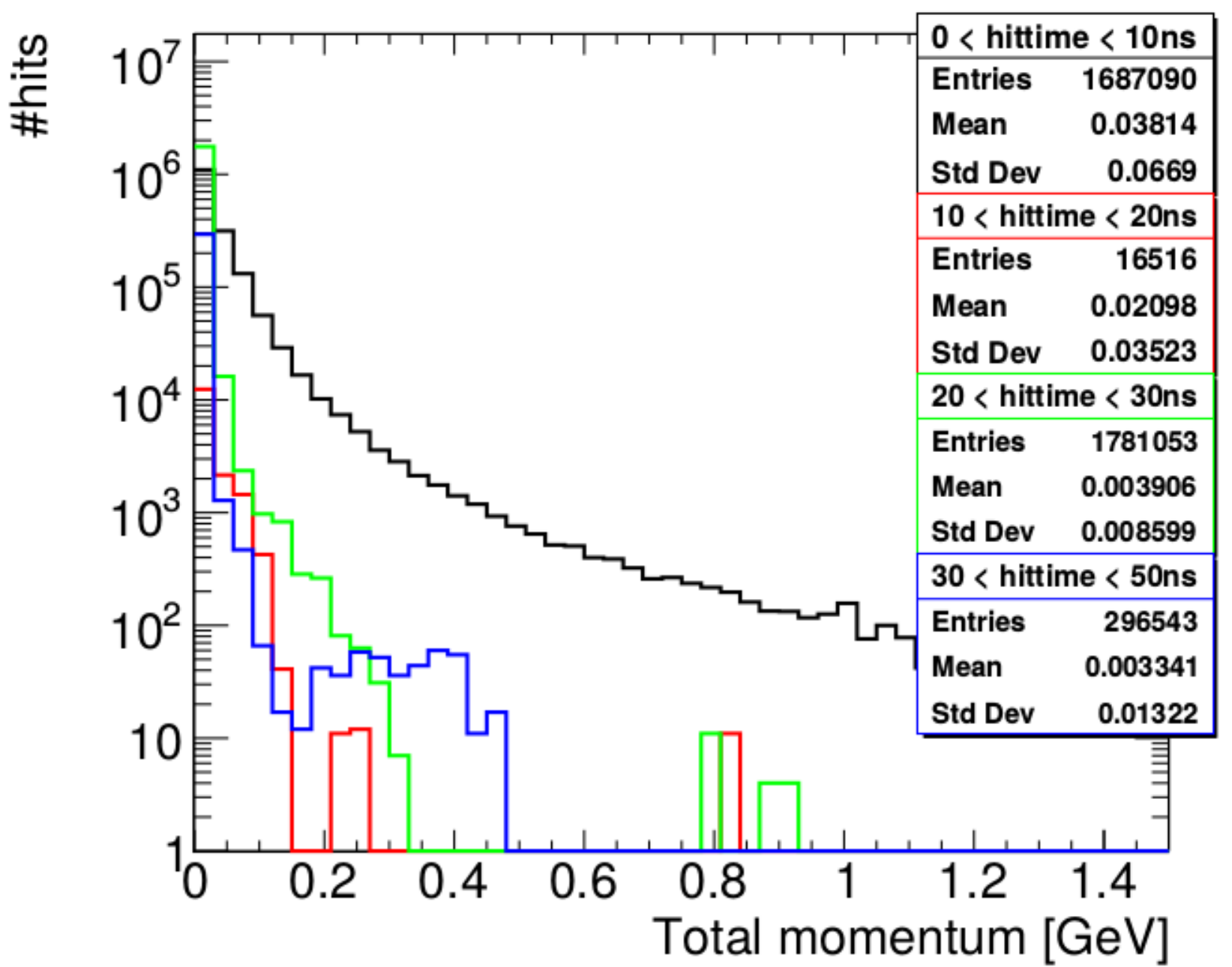}
\end{subfigure}
\begin{subfigure}[t]{0.5\textwidth}
\centering
\includegraphics[width=\textwidth]{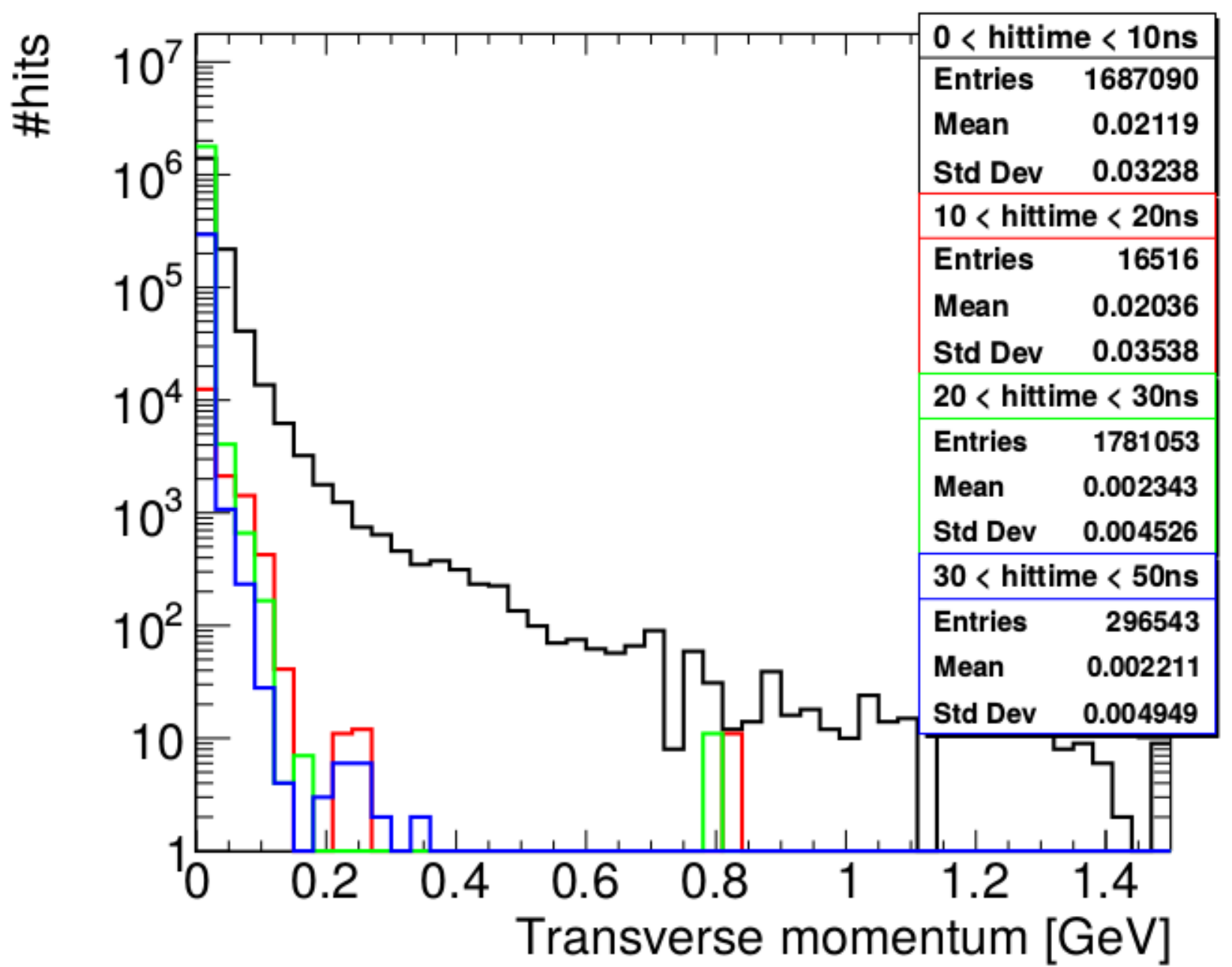}
\end{subfigure}
\caption[Momentum distribution of pair background particles hitting the barrel and the endcaps of the SiD vertex detector]{Distributions of the pair background particle momenta of pair background particles that will hit the barrel and the endcaps of the vertex detector. For higher statistics, 1312 bunches are accumulated. The plots show the histograms of the total and the transverse momenta of the particles hitting the vertex detector, in certain time intervals: $]\unit[0]{ns};\unit[10]{ns}]$,$]\unit[10]{ns};\unit[20]{ns}]$,$]\unit[20]{ns};\unit[30]{ns}]$ and $]\unit[30]{ns};\unit[50]{ns}]$.}
\label{fig:momentum_VtxEndcapBarrel}
\end{figure}

\begin{figure}
\centering
\includegraphics[width=0.85\textwidth]{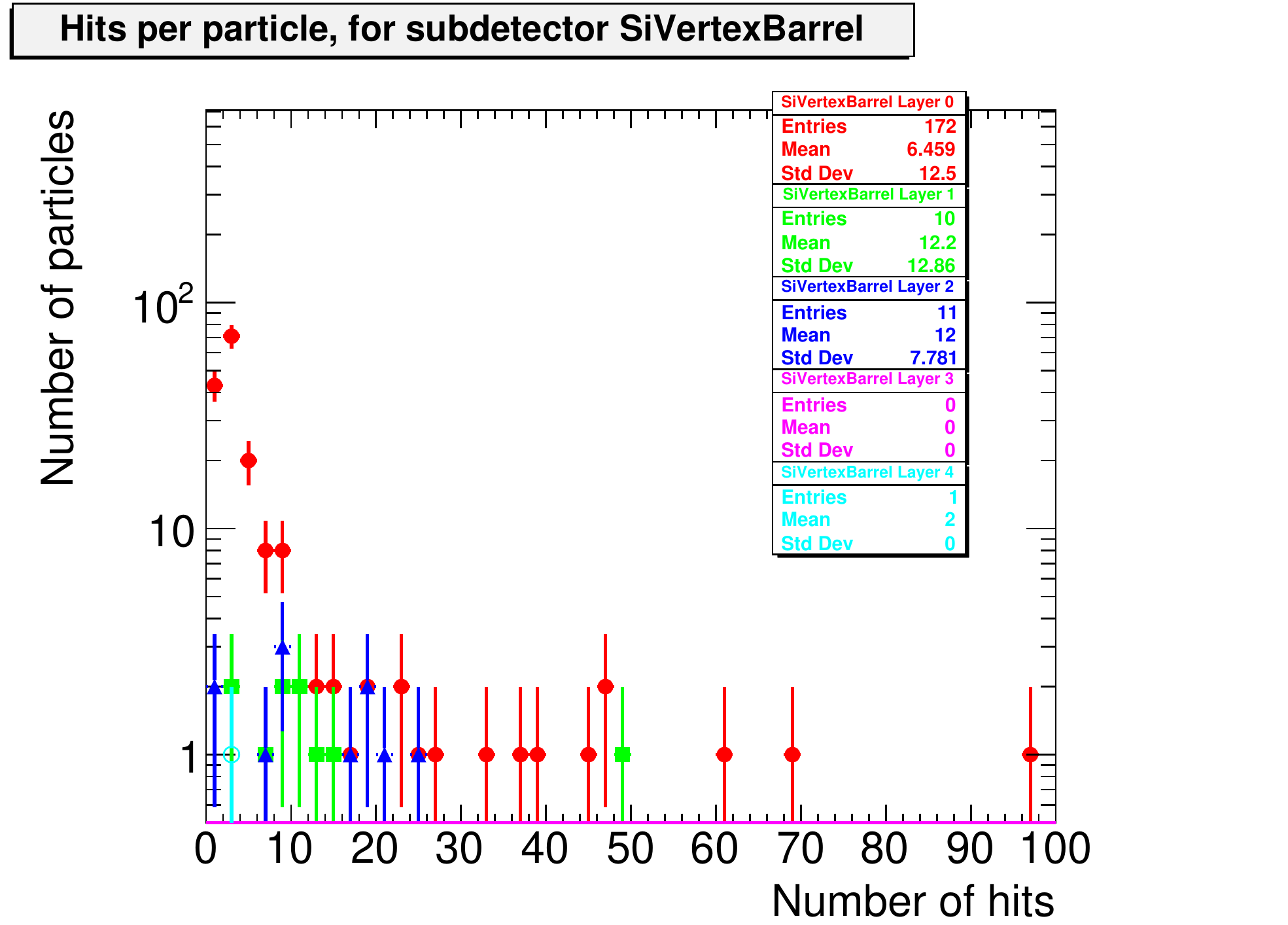}
\caption[Number of hits per pair background particle hitting the barrel of the SiD vertex detector]{The number of hits that the pair background particles from a single bunch produce in the vertex barrel by spiraling in the solenoid field due to their low transverse momentum. The plot shows the number of pair  background particles that hit the vertex barrel a certain number of times. In the innermost layer, there are particles hitting the vertex detector even between 60 and 100 times because of spiraling in the magnetic solenoid field.}
\label{fig:HitsPerParticle_VtxBarrel}
\end{figure}

With large samples it is clear that the bulk of pair background particles originate from the center of the detector at the time of the bunch crossing. There are also hits from particles originating from the detector center part arriving in the gap between 10 and \unit[20]{ns}. By looking closely at the particles originating from the interaction point, one can see that for all hit time intervals a large number of particles originate from there. These are primary particles that are created at the time of the bunch crossing but arrive at the vertex detector several nanoseconds to several microseconds later. The reason for this is their low transverse momentum of only up to approximately \unit[0.2]{GeV}, which is plotted in Figure~\ref{fig:PT_hittime_VtxEndcapBarrel}.
Additionally, histograms were made for both the total and the transverse momentum of the primary pair background particles and the backscatter pairs, shown in Figure~\ref{fig:momentum_VtxEndcapBarrel}. Here, as well as for several figures before, the histograms were made for the four time intervals ]\unit[0]{ns};\unit[10]{ns}], ]\unit[10]{ns};\unit[20]{ns}], ]\unit[20]{ns};\unit[30]{ns}] and ]\unit[30]{ns};\unit[50]{ns}]. As expected, the momentum distribution of the pair background particles primarily produced at the time of the bunch crossing is rather broad, while the total momenta of the backscatter particles only reach up to $\sim$\unit[0.4]{GeV} and the transverse momenta only up to $\sim$\unit[0.2]{GeV} as shown before.
These low $p\_\text{T}$ particles are deflected by the solenoid field onto a spiral track with a radius of up to about \unit[5]{cm}. However, the majority of the late-arriving particles spiral with a radius less than \unit[1]{cm}. The large amount of hits in the vertex detector from particles originating from the interaction point is therefore due to the fact that the particles are looping through the detector and leaving up to around 100 hits in the vertex detector, which can be seen in Figure~\ref{fig:HitsPerParticle_VtxBarrel}.
\\To summarize, of all primary pair background particles that hit the vertex detector, about 68\% have such a low $p_\text{T}$ that they therefore spiral and produce a large amount of hits in the vertex detector later than \unit[1]{ns} after the bunch crossing.

\begin{figure}
\centering
\includegraphics[width=0.8\textwidth]{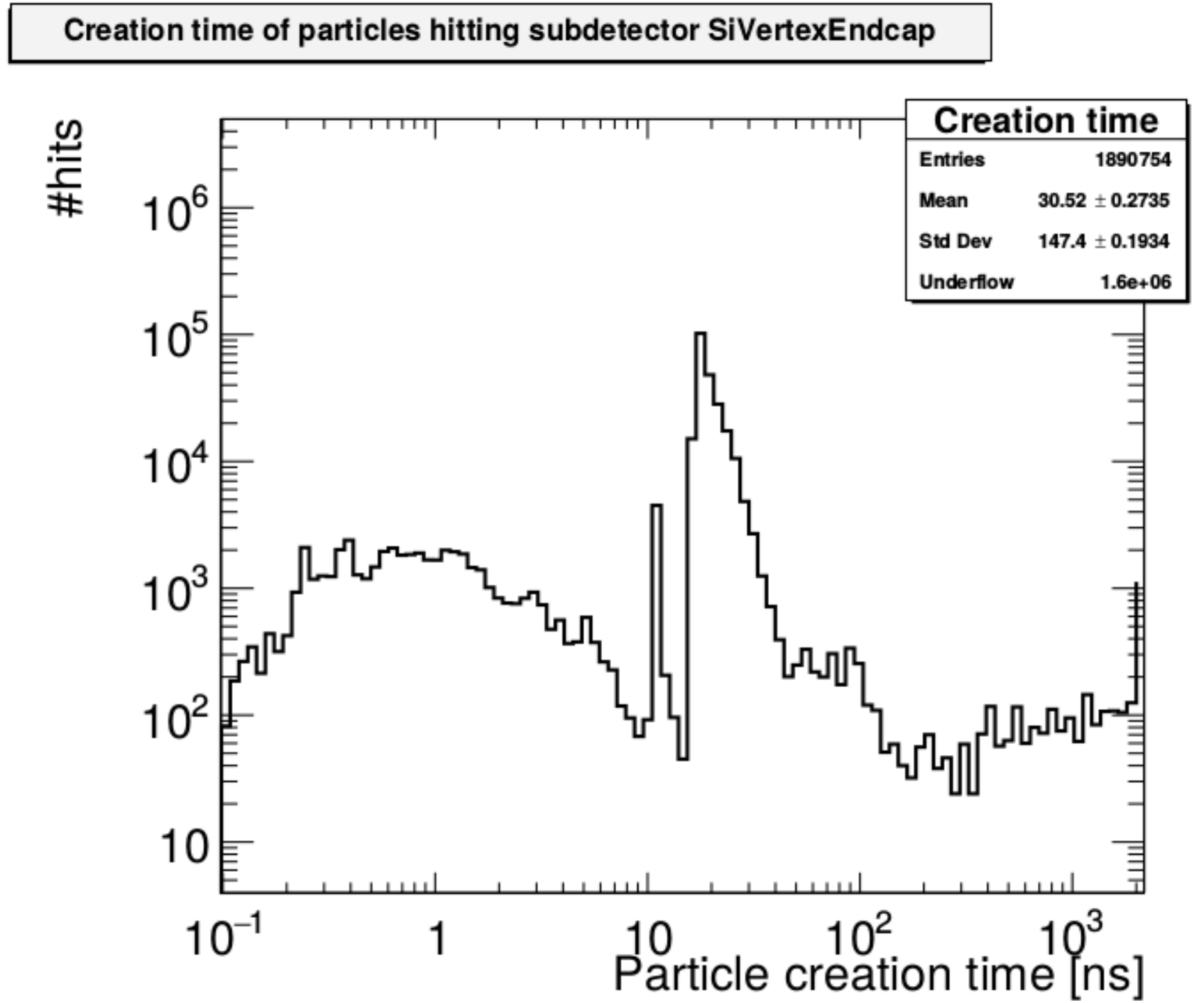}
\caption[Creation time distribution of pair background particles hitting the endcaps of the SiD vertex detector]{Distribution of the time of creation (relative to the instant of the bunch crossing at \unit[0]{ns}) of pair background particles that hit the endcaps of the vertex detector. Again, the distributions are for 1312 bunches. At the time of the bunch crossing about \num{1.6e6} particles are created. This number is shown in the underflow bin.\\To be noted is that the creation time here is plotted for pair background particles hitting the vertex endcaps only. This is to avoid double counting of particles that would hit both, the barrel and the endcaps. The shape of the distribution does not change significantly.}
\label{fig:creationtime_VtxEndcapBarrel}
\end{figure}

\begin{figure}
\begin{subfigure}[t]{0.5\textwidth}
\centering
\includegraphics[width=\textwidth]{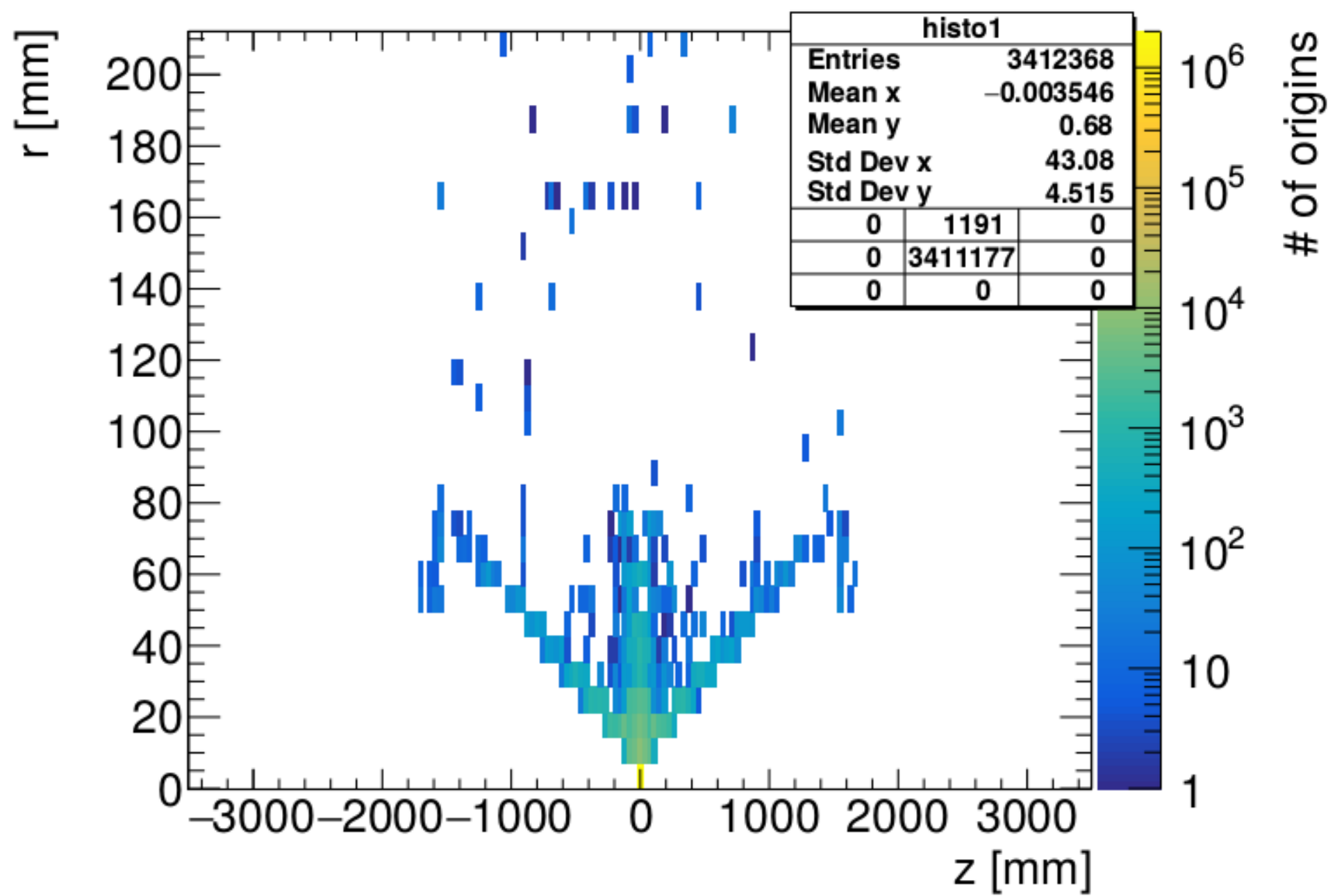}
\caption{Created in $[\unit[0]{ns};\unit[10]{ns}]$}
\end{subfigure}
\begin{subfigure}[t]{0.5\textwidth}
\centering
\includegraphics[width=\textwidth]{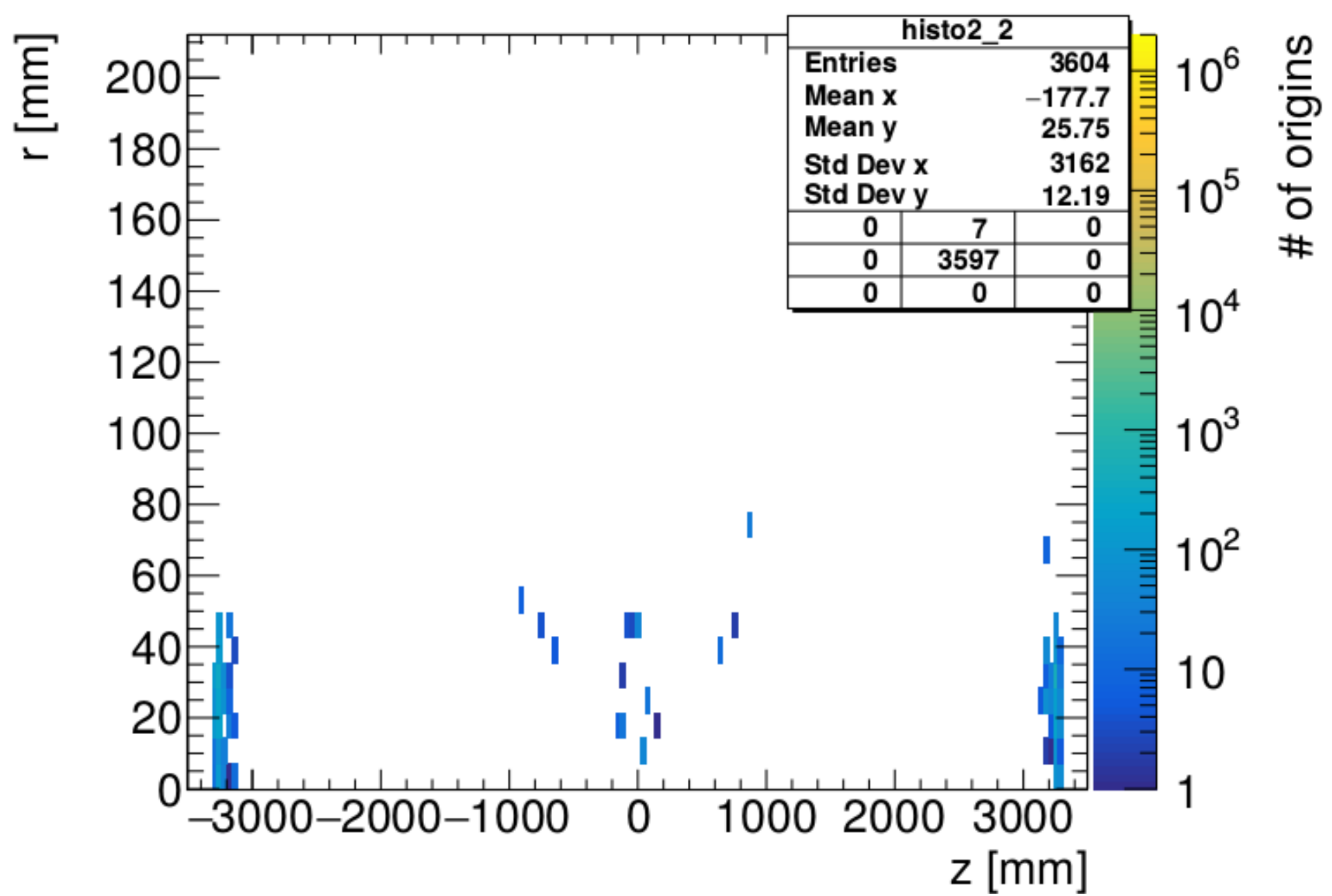}
\caption{Created in $]\unit[10]{ns};\unit[11]{ns}]$}
\end{subfigure}
\begin{subfigure}[t]{0.5\textwidth}
\centering
\includegraphics[width=\textwidth]{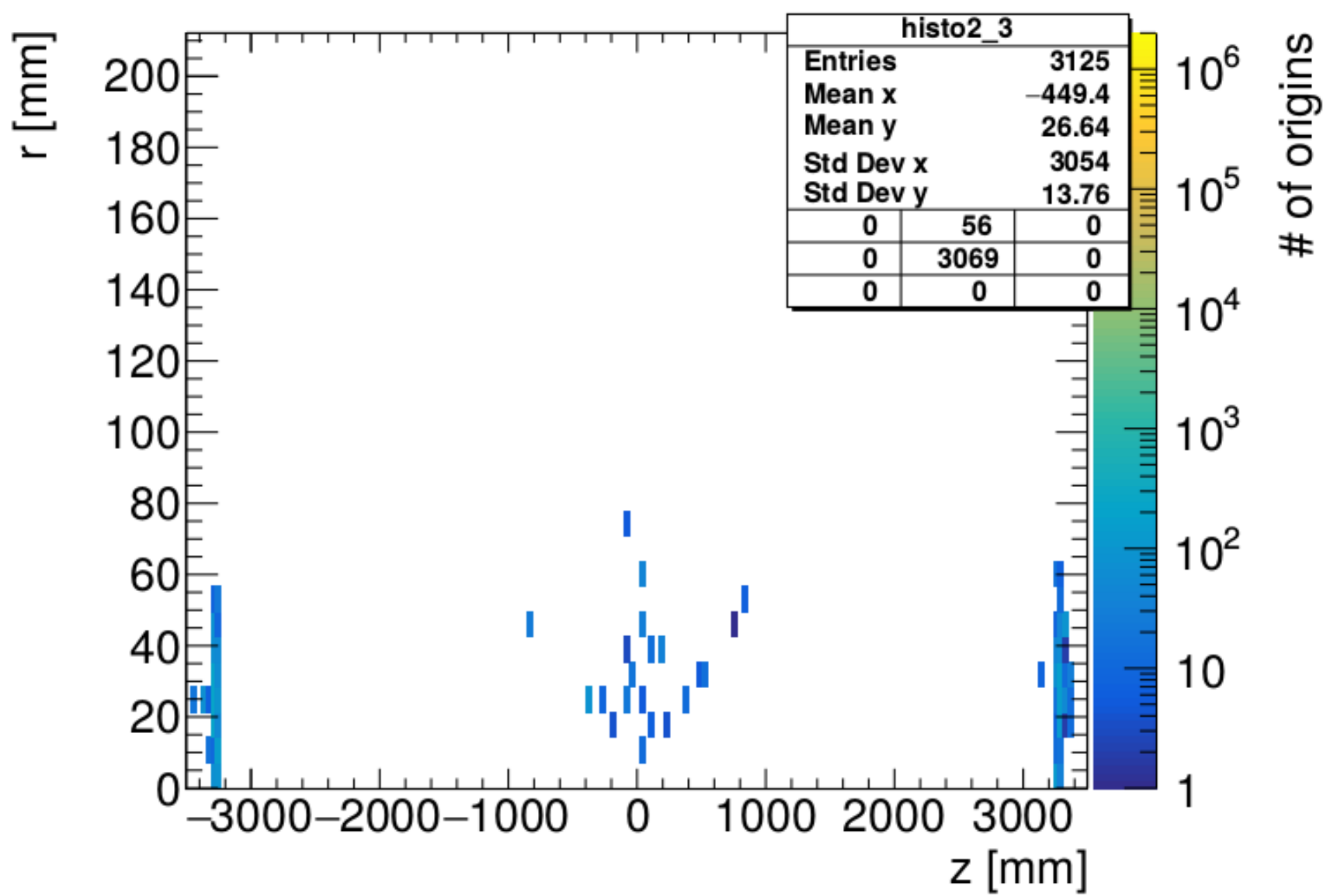}
\caption{Created in $]\unit[11]{ns};\unit[13]{ns}]$}
\end{subfigure}
\begin{subfigure}[t]{0.5\textwidth}
\centering
\includegraphics[width=\textwidth]{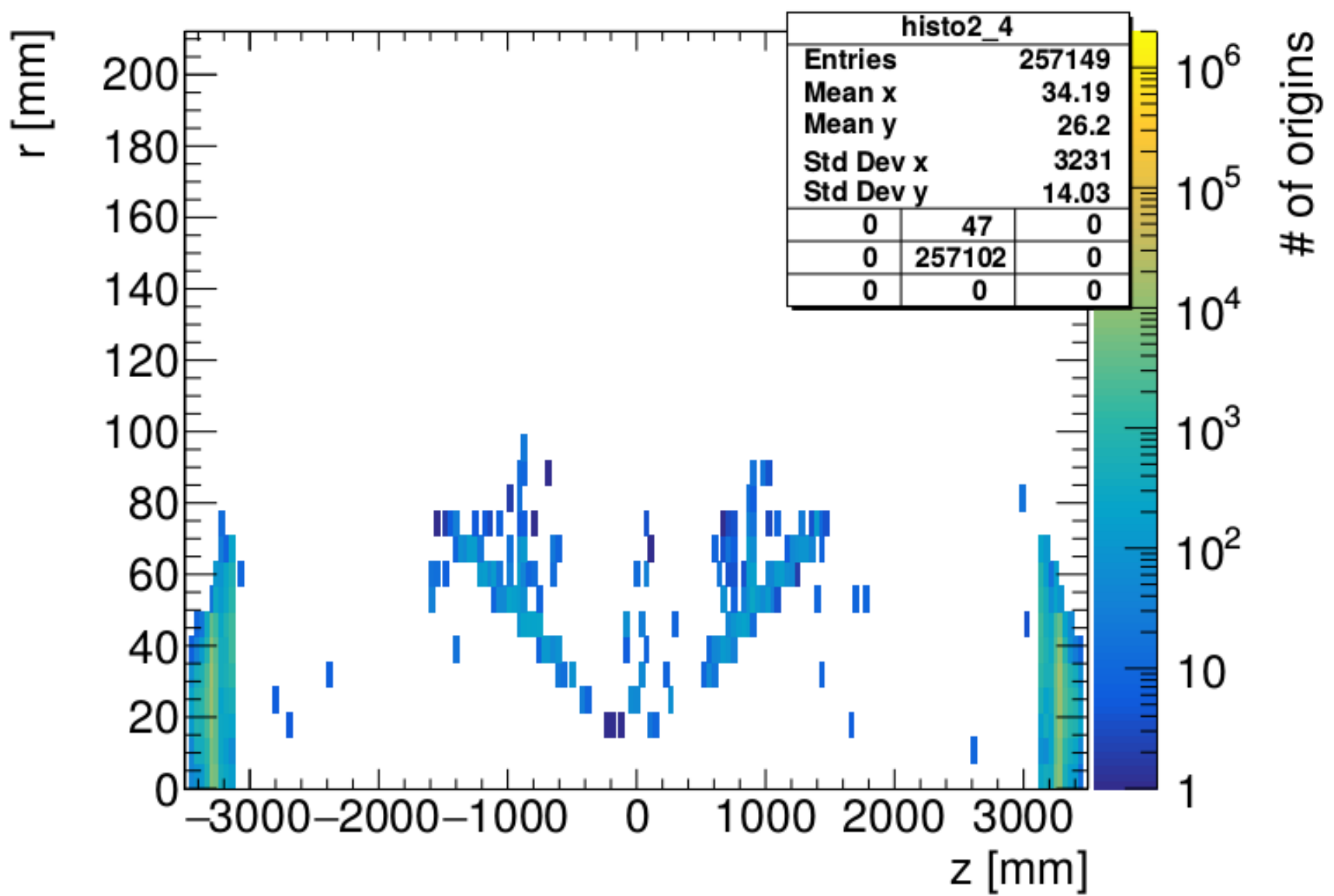}
\caption{Created in $]\unit[13]{ns};\unit[20]{ns}]$}
\end{subfigure}
\begin{subfigure}[t]{0.5\textwidth}
\centering
\includegraphics[width=\textwidth]{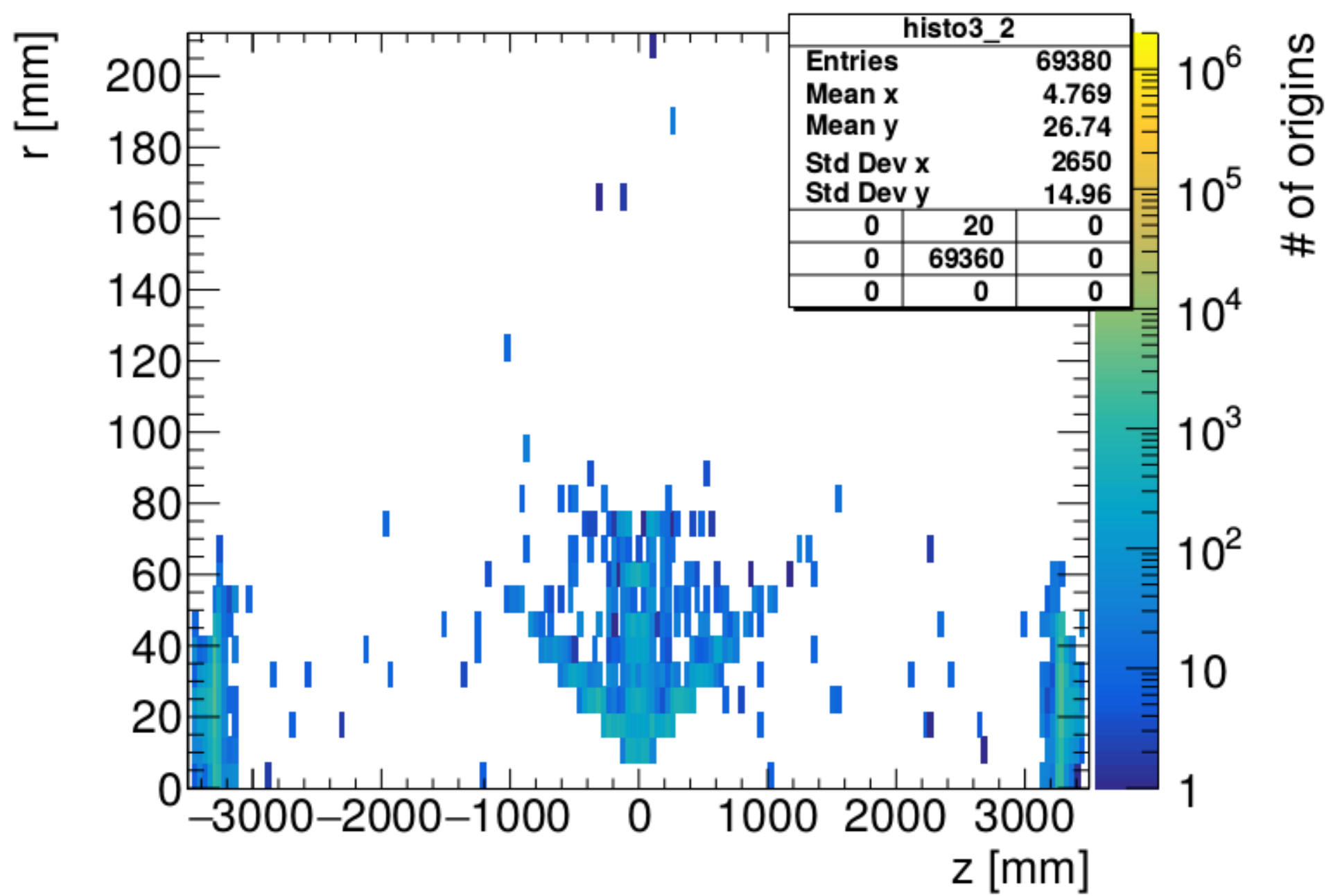}
\caption{Created in $]\unit[20]{ns};\unit[23]{ns}]$}
\end{subfigure}
\begin{subfigure}[t]{0.5\textwidth}
\centering
\includegraphics[width=\textwidth]{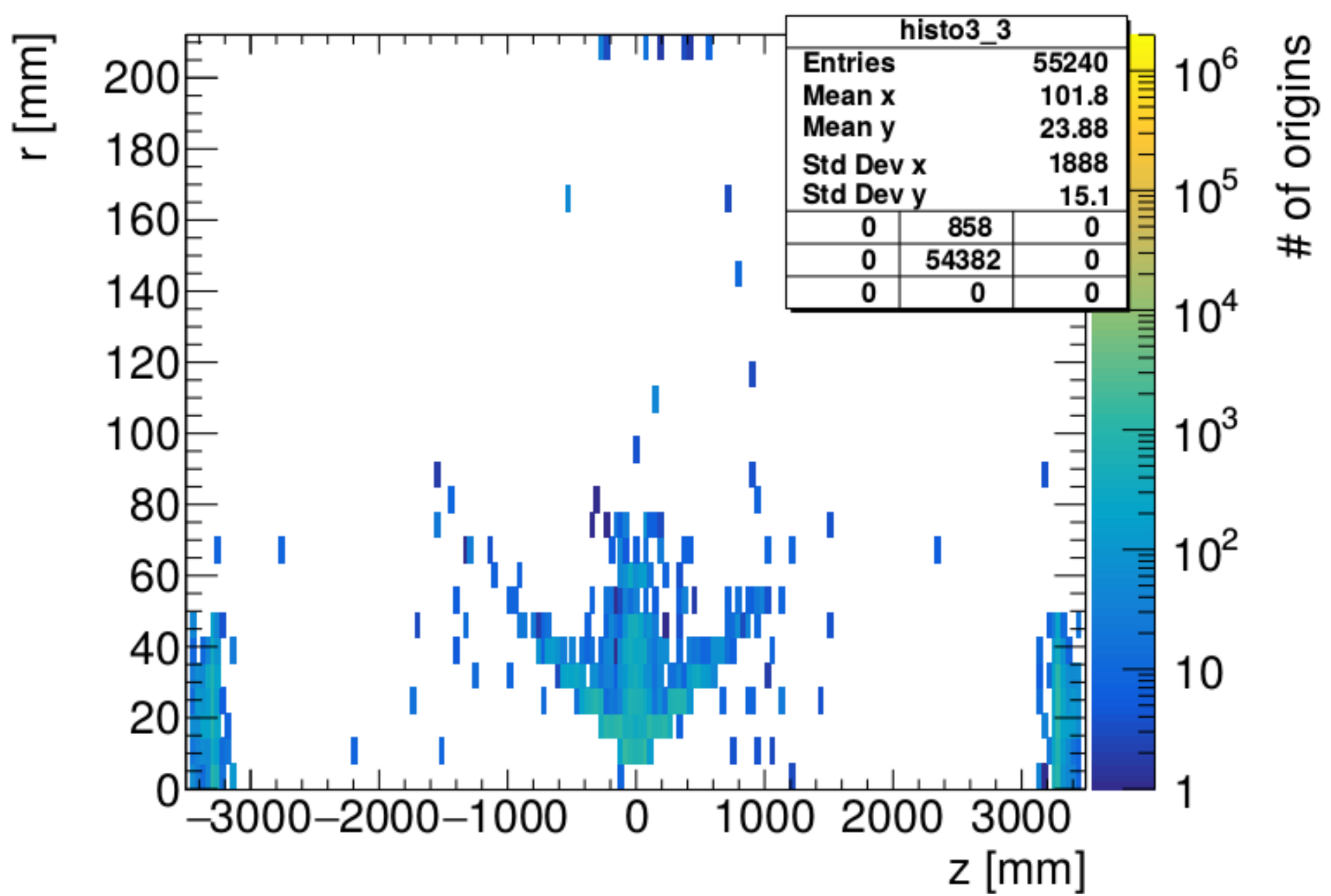}
\caption{Created in $]\unit[23]{ns};\unit[30]{ns}]$}
\end{subfigure}
\begin{subfigure}[t]{0.5\textwidth}
\centering
\includegraphics[width=\textwidth]{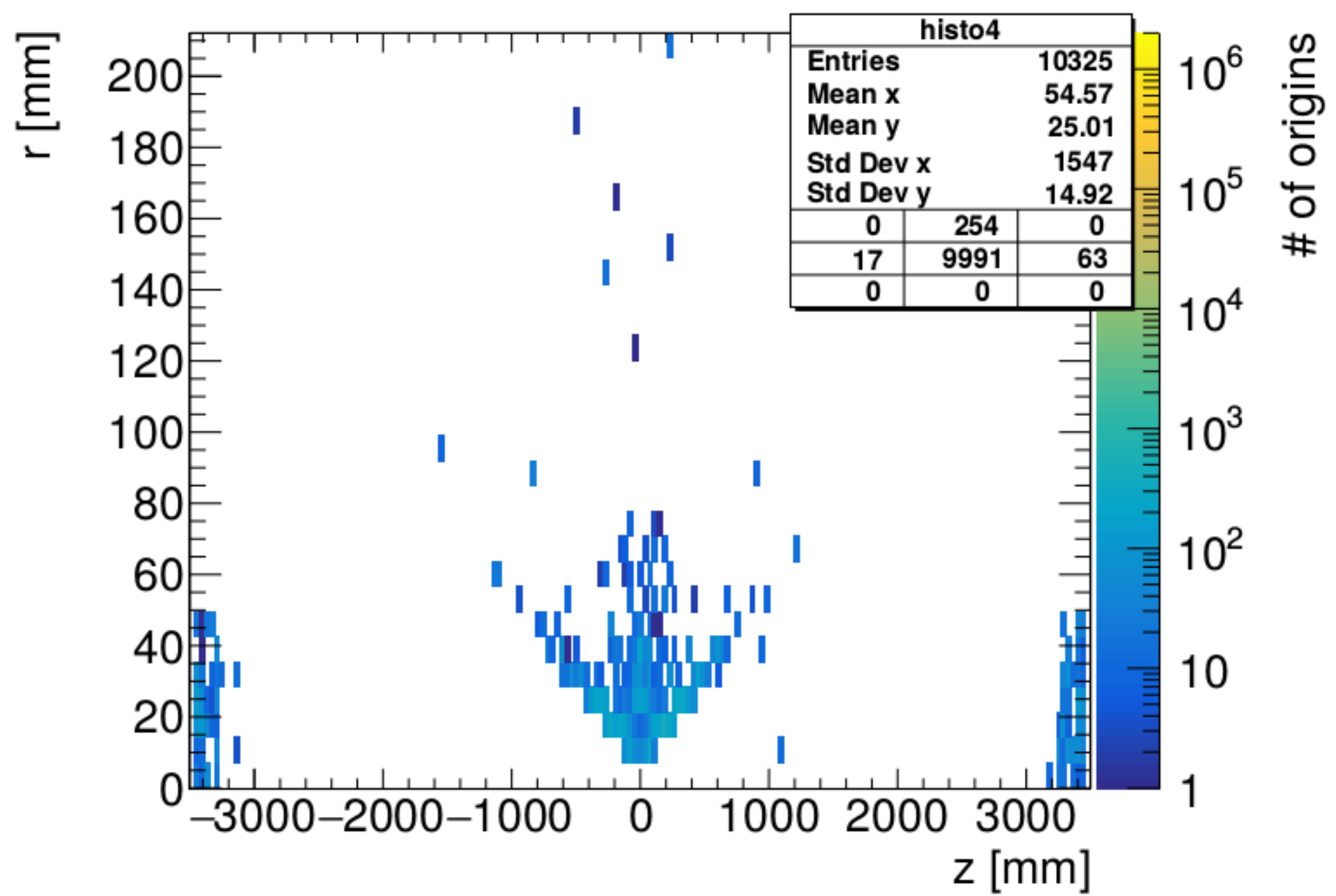}
\caption{Created in $]\unit[30]{ns};\unit[50]{ns}]$}
\end{subfigure}
\begin{subfigure}[t]{0.5\textwidth}
\centering
\includegraphics[width=\textwidth]{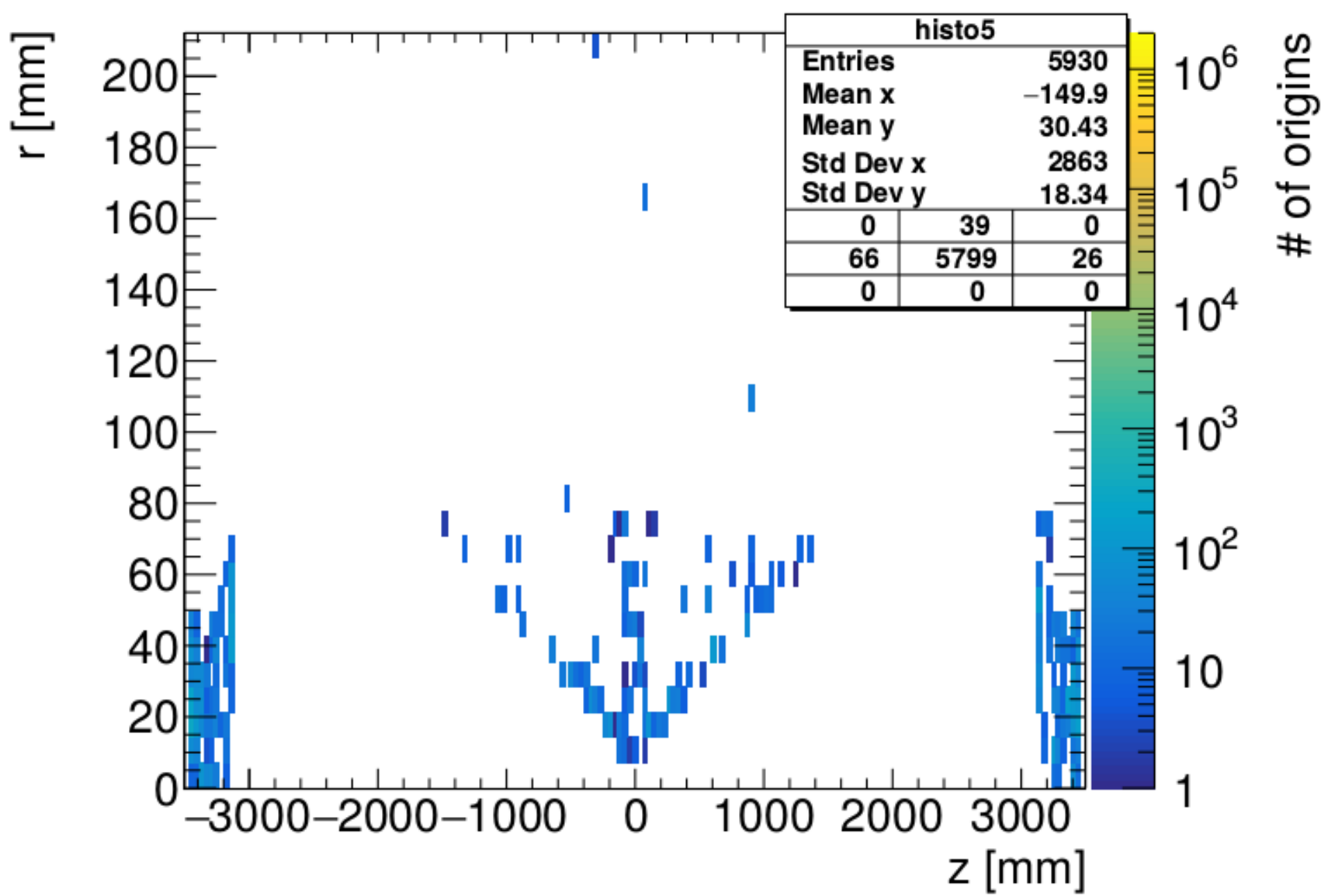}
\caption{Created in $]\unit[50]{ns};\unit[1000]{ns}]$}
\end{subfigure}
\caption[Origins of pair background particles hitting the barrel and the endcaps of the SiD vertex detector]{Point of origin of pair background particles that hit the barrel and the endcaps of the vertex detector, in slices of the time of creation of the background particles. To increase the statistics, 1312 bunches are plotted.}
\label{fig:particleorigin_vertexendcap_creationtime}
\end{figure}

The source of the hit time discrimination can be further understood by looking at the time of creation of these pair background particles, as shown in Figure~\ref{fig:creationtime_VtxEndcapBarrel}. To further illuminate the distribution of Figure~\ref{fig:creationtime_VtxEndcapBarrel}, the points of origin of the background particles are plotted again in Figure~\ref{fig:particleorigin_vertexendcap_creationtime}, but now in slices of the time of the creation of the background particles, not in slices of the hit time as in Figure~\ref{fig:particleorigin_vertexendcap}. Figures~\ref{fig:creationtime_VtxEndcapBarrel}
and~\ref{fig:particleorigin_vertexendcap_creationtime} can be directly compared:
As expected, the first peak in the creation time plot drops slowly, where the pair background particles are originating from the interaction region. The peaks between between 13 and \unit[50]{ns} are mirrored in the subfigures of the particle origin plots in Figure~\ref{fig:particleorigin_vertexendcap_creationtime}, where it is visible that particles are being reflected from the endcaps of the SiD detector, i.e. the BeamCal and the flux return. The time between this peak at $\sim$\unit[13]{ns} and the bunch crossing time is therefore the time of flight between the IP and the SiD endcaps. The long tail of out-of-time backscatter particles from \unit[50]{ns} on shows that particles can arrive at the vertex detector up to several microseconds late, i.e. later than the next bunch crossing. The percentages of particles created in these certain time intervals are listed in Table~\ref{tab:Percentages_Pairs_CreationTime}. With a time gab of \unit[554]{ns} between two bunches, about \num{2.9E-4} of all pair background particles hitting the vertex detector endcaps will overlay with the following bunch. The total number of overlaying particles increases when looking at all the particles, not only those hitting the vertex detector.\\
Looking back at Figure~\ref{fig:particleorigin_vertexendcap}, one can see the same structures in the distributions of the creation time and the particle origins in terms of the creation time. As the particle time-of-flight takes up a certain time, the peaks of the creation-time plots are to be seen several nanoseconds later in Figure~\ref{fig:particleorigin_vertexendcap}.
With this knowledge, the time distributions of hits of single bunches in Figures~\ref{fig:hittime_VtxBarrel_1bunch} and \ref{fig:hittime_VtxEndcap_1bunch} can now be explained easily.

\begin{table}
\caption{Particle types shown as a percentage of the total number of particles hitting the vertex detector endcaps. The total number for a full train of bunches is 1,890,754 as can be seen in Figure~\ref{fig:creationtime_VtxEndcapBarrel}. The pair background particles are created in certain time intervals: in the time interval [\unit[0]{ns};\unit[11]{ns}] the primary pairs are created, late secondaries are created in the interval ]\unit[11]{ns};\unit[50]{ns}]. Out-of-time backscatter pairs are defined as those pairs that are created later than \unit[50]{ns} after the bunch crossing. The percentages are calculated from the number of entries in the certain time intervals in Figure~\ref{fig:creationtime_VtxEndcapBarrel}.}
\label{tab:Percentages_Pairs_CreationTime}
\begin{tabular}{>{\RaggedRight}p{3cm}>{\RaggedRight}p{3cm}>{\RaggedRight}p{3.1cm}>{\RaggedRight}p{3.1cm}}
\hline\hline
Primary pairs [\unit[0]{ns};\unit[11]{ns}] & Late pairs ]\unit[11]{ns};\unit[50]{ns}] & Out-of-time backscatter pairs ]\unit[50]{ns};\unit[554]{ns}] & Out-of-time backscatter pairs ]\unit[554]{ns};\unit[1000]{ns}]\\
\hline
87.33\% & 12.38\% & 0.16\% &  0.029\% \\
\hline\hline
\end{tabular}
\end{table}

In conclusion, the time distribution studies can give a better understanding of the hit time of the background particles and the effect of backscatter particles on the detector occupancy. Depending on the time gap between the bunch crossing and the arrival time of these backscatter particles, there is an opportunity to cut away background events by digitally/electronically applying time gates. On the other hand, very slow backscatter particles that overlay with the background particles and the primary particles from the next bunch can lead to higher occupancies in the subdetectors and an increase in underlying events. This effect, however, is small with e.g. only \num{2.9E-4} of all particles hitting the vertex endcap later than \unit[554]{ns} after the first bunch crossing.


\section{BeamCal Reconstruction}

While, as discussed above, the albedo formed by the interaction of the pair background with the BeamCal can compromise the performance of the inner tracking layers of the SiD detector, the pairs also form a background to the detection of individual high-energy electrons and positrons in the BeamCal itself. These backgrounds are particularly problematic for the detection of intermediate-energy electrons (50 GeV and below) in the regions of the BeamCal closest to the incoming and outgoing beams. Since the detection of these electrons can be important to the study of degenerate SUSY scenarios (a typical scenario would be stau pair production, with corresponding decays to $\tau + \chi_1^0$, with the $\chi_1^0$ close in mass to the stau), it is important to understand how the various ILC and BeamCal configurations discussed above will impact the BeamCal reconstruction efficiency.

To search for individual high-energy electrons and positrons in the BeamCal, the 10th layer (close to the shower maximum for the range of energies of interest) is used as a seed layer. The mean pair background is subtracted from each channel, and then the 50 highest-energy channels are extended back to layer 40, adding the energy in the corresponding channel in each of those thirty deeper layers to form a set of 50 ``cylinder'' energy sums, each of which is a potential electron/positron candidate. Each of these 50 candidate cylinder energy sums is compared to the background arising from pairs in the given cylinder, and if the energy is above a pre-determined number of standard deviations of the pair background energy distribution, the cylinder is identified as an electron or positron candidate. The number-of-standard-deviations requirement is the same across the face of the BeamCal (i.e., is independent of the cylinder under consideration), and is selected so that exactly 10\% of beam bunch train for which only pairs reach the BeamCal are mistakenly identified as having an electron or positron candidate.

\begin{figure}
\centering
\includegraphics[width=0.8\textwidth]{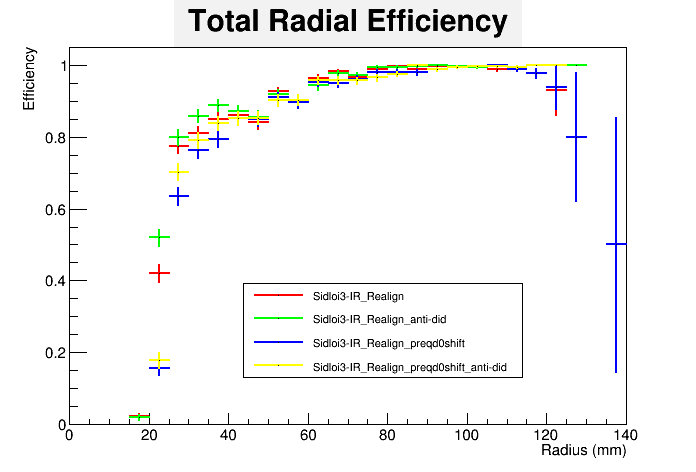}
\caption[Beamstrahlung processes]{Radial dependence of the total BeamCal reconstruction efficiency. In the key, ``preqd0shift'' refers to a configuration with an L* of 3.5 m as opposed to the nominal L* of 4.1 m. The designation ``anti-did'' refers to the inclusion of the anti-DiD field; otherwise, no such field is present in the simulation.}
\label{fig:eff_tot}
\end{figure}

\begin{figure}
\centering
\includegraphics[width=0.8\textwidth]{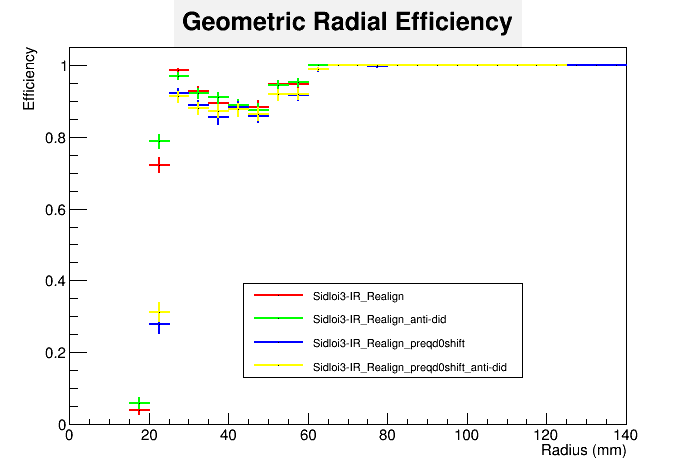}
\caption[Beamstrahlung processes]{Radial dependence of the geometrical BeamCal reconstruction efficiency. In the key, ``preqd0shift'' refers to a configuration with an L* of 3.5 m as opposed to the nominal L* of 4.1 m. The designation ``anti-did'' refers to the inclusion of the anti-DiD field; otherwise, no such field is present in the simulation.}
\label{fig:eff_geo}
\end{figure}

\begin{figure}
\centering
\includegraphics[width=0.8\textwidth]{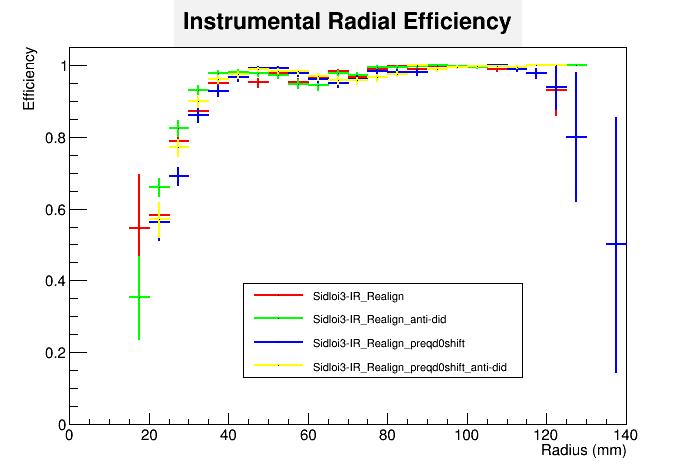}
\caption[Beamstrahlung processes]{Radial dependence of the instrumental BeamCal reconstruction efficiency. In the key, ``preqd0shift'' refers to a configuration with an L* of 3.5 m as opposed to the nominal L* of 4.1 m. The designation ``anti-did'' refers to the inclusion of the anti-DiD field; otherwise, no such field is present in the simulation.}
\label{fig:eff_inst}
\end{figure}

Figure~\ref{fig:eff_tot} shows the resulting (total) BeamCal reconstruction efficiency for 50 GeV electrons as a function of radius from the center of the outgoing beampipe, for the different ILC and BeamCal configurations discussed in the previous sections. Figures~\ref{fig:eff_geo} and~\ref{fig:eff_inst} show the factorization of the total efficiency of Figure~\ref{fig:eff_tot} into its "geometric" (Figure~\ref{fig:eff_geo}) and "instrumental" (Figure~\ref{fig:eff_inst}) components. For the geometric component, any electron that strikes the active area of the BeamCal is included in the numerator of the efficiency quotient, regardless of whether a candidate is identified for that bunch train according to the criteria of the preceding paragraph. For the instrumental efficiency, only electrons for which a candidate is identified for the given bunch train are included in the efficiency numerator, but in addition, only electrons that strike the active area of the BeamCal are included in the numerator. In this way, the geometrical efficiency relates to the geometrical coverage of the BeamCal instrument, while the instrumental efficiency relates to the reconstruction capability of the BeamCal design, given the pair backgrounds associated with the given ILC/BeamCal configuration. The total BeamCal reconstruction efficiency is, by construction, the product of the geometrical and instrumental efficiencies.

Comparing the red and blue trajectories in Figure~\ref{fig:eff_tot}, it is seen that the overall BeamCal reconstruction efficiency improves with the increase of L* from 3.5 m to 4.1 m. Examining the same trajectories in Figures~\ref{fig:eff_geo} and~\ref{fig:eff_inst} suggests that the difference is primarily geometrical: with the longer lever-arm associated with the larger value of L*, more electrons at smaller angles strike the instrumented region of the BeamCal, whose geometry is independent of L*. 

\begin{figure}
\centering
\includegraphics[width=0.8\textwidth]{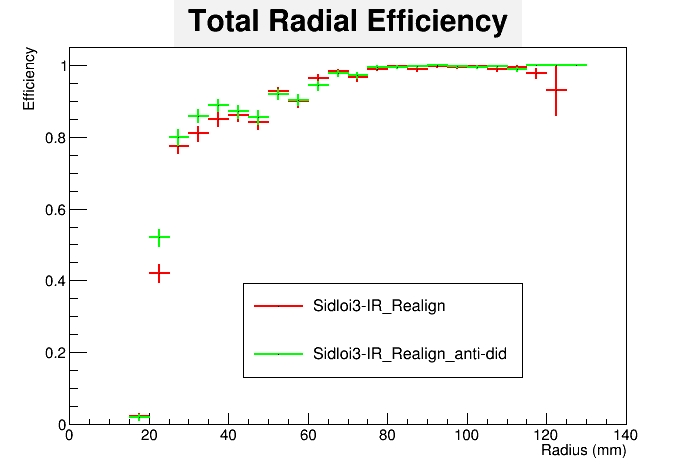}
\caption[Beamstrahlung processes]{Effect of the anti-DiD field upon the total efficiency of the BeamCal reconstruction, as a function of the radius of the incident 50 GeV electron or positron.}
\label{fig:eff_did}
\end{figure}

\begin{figure}
\centering
\includegraphics[width=0.8\textwidth]{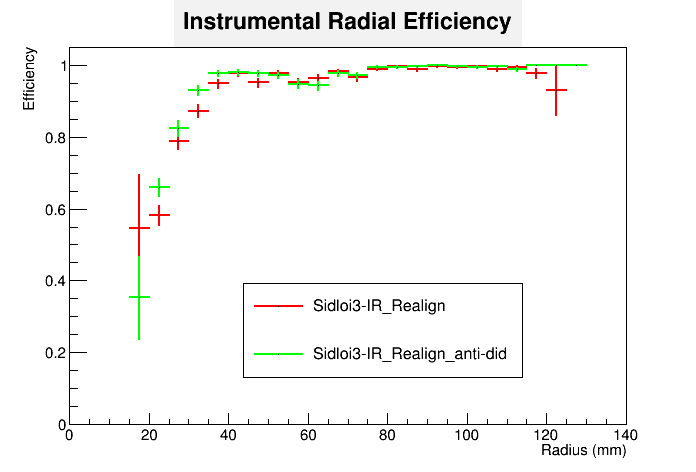}
\caption[Beamstrahlung processes]{Effect of the anti-DiD field upon the instrumental component of the BeamCal reconstruction efficiency, as a function of the radius of the incident 50 GeV electron or positron.}
\label{fig:eff_did_inst}
\end{figure}

Comparing the red and green trajectories, and the blue and yellow trajectories, in these figures one can look for the effect of the anti-DiD upon the BeamCal reconstruction efficiency. For the purpose of clarity, the effect of the anti-DiD field on the total BeamCal efficiency is shown directly in Figure~\ref{fig:eff_did} for the case of L* = 4.1 m; since the geometry is not affected by the inclusion of the anti-DiD field, the observed effect is purely instrumental. A noticeable improvement in the efficiency is observed for radii below 40~mm. Although the effect may appear to be small, it should be borne in mind that a small decrease in the BeamCal reconstruction efficiency can result in a large increase in un-vetoed two-photon backgrounds, in the case that the instrumental component of the BeamCal efficiency approaches unity when the anti-DiD is included. In this vein, Figure~\ref{fig:eff_did_inst} shows the instrumental component of the BeamCal reconstruction efficiency, with and without the anti-DiD field. For radii between 30 and 40 mm, the BeamCal inefficiency for 50 GeV electrons seems to be significantly reduced by the inclusion of the anti-DiD field. Dedicated physics studies, such as the reach in degenerate SUSY scenarios, will need to be undertaken in order to assess the value of the advantage afforded by the inclusion of the anti-DiD.

\section{Summary and Conclusions}

Making use of the SiD simulation infrastructure, we have conducted a simulation study of several aspects of the performance of the SiD vertex detector, forward electromagnetic calorimeter, and BeamCal. In particular, we have explored the way in which collision backgrounds (primarily from the production of coherent electron-positron pairs, but also including two-photon events when necessary) impact the performance of these detector elements. Within this context, we have also explored the effect of various IP design choices, including L* and BeamCal geometry variants as well as the inclusion of an anti-DiD magnetic field configuration that would sweep much of the coherent pair background into the exiting beampipe, thereby potentially improving the BeamCal reconstruction efficiency and lowering vertex detector backgrounds.

Fractional channel occupancy was used as a measure of the impact of the pair background on the performance of the vertex detector. Even under the conservative assumption of a pixel size of $\unit[30\times30]{\micron^{2}}$ and a readout that integrates over five bunch crossings, the occupancy was found to be well below $10^{-3}$ for any point in the barrel portion, and for any point in the endcap except at the innermost radii, where it rose to approximately $2 \times 10^{-3}$. Studies of the point-of-origin of pair-induced vertex detector background showed that only a fraction of the background arises from the BeamCal albedo, making the result insensitive to the L*, anti-DiD, and BeamCal geometry variants that were studied.

Furthermore, the studies of the hit time and creation time distributions of the background particles have shown that the pairs from the pair background do not hit the vertex detector instantaneously at the time of the bunch crossing. Due to the effect of backscattering and their low transverse momentum, pair background particles hit the vertex detector up to several microseconds late. These broad time distributions will allow us to reject hits that occur more than \unit[20]{ns} after the instant of the beam crossing by use of an electronic gate.

Making use of a simulation of all potentially high-occupancy backgrounds (pair, Bhabha, and two-photon production), we have explored the necessary buffer depth of the forward electromagnetic calorimeter readout. For the forward calorimeter overall, a depth of four buffers leads to a loss of several tenths of a percent of hits. Increasing this to six reduces the hit loss to just above  $10^{-4}$. However, while this result was also observed when the layer-by-layer hit loss was evaluated, the inverse relationship between background rate and distance from the beampipe led to a requirement of eight buffers to maintain hit losses below $10^{-4}$ for the innermost radii of the forward electromagnetic calorimeter.

Finally, our study of the dependence of the BeamCal reconstruction efficiency upon the choice of L* showed little effect, with a slight improvement in the geometrical acceptance observed for the larger, preferred value of L*. The inclusion of the anti-DiD field, which can have no effect on the geometrical acceptance, was observed to provide a noticeable improvement in the inefficiency for tagging \unit[50]{GeV} electrons in the region of the BeamCal between 30 and \unit[40]{cm} from the exiting electron/positron beams.



\section*{Acknowledgments}
A portion of the research was performed using PNNL Institutional Computing at the Pacific Northwest National Laboratory. This work was in part supported by the Nuclear Physics, Particle Physics, Astrophysics and Cosmology Initiative, a Laboratory Directed Research and Development program at the Pacific Northwest National Laboratory. \\
The Santa Cruz group gratefully acknowledges the support of the Department of Energy under grant number DE-SC0010107.\\
At the DESY National Analysis Facility NAF2 and its batch system BIRD, the complete set of the pair background events from beam-beam interactions was generated, which is used for the vast majority of the background studies.

\section{Appendix}
\subsection{GuineaPig acc.dat}
\label{sec:Appendix_GuineaPig}
The nominal parameters used as input for GuineaPig are given in the following. The pair background, that was analyzed for this note, was generated with these GuineaPig parameters:\\
\texttt{\$ACCELERATOR:: ILC-500GeV\\
\{energy=250.0; particles=2.0; beta\_x=11.0; beta\_y=0.48;\\
emitt\_x=10.0; emitt\_y=0.035; sigma\_z=300.0; f\_rep=5.0; n\_b=1312;\\
charge\_sign=-1; scale\_step=1.0; waist\_y=250;\\
espread.1=0.00124; espread.2=0.0007; which\_espread=3;\}\\
\$PARAMETERS:: par\_pairs\\
\{n\_z=12; n\_t=6; n\_m=80000; cut\_z=3.5*sigma\_z.1; n\_x=256; n\_y=256;\\
cut\_x=4*sigma\_x.1; cut\_y=4*sigma\_y.1; pair\_ecut=1e-3; pair\_q2=2;\\
beam\_size=1; grids=7; store\_beam=1; do\_pairs=1; track\_pairs=1;\\
store\_pairs=1; do\_photons=1; store\_photons=1; do\_hadrons=1;\\
do\_jets=1; do\_coherent=1; electron\_ratio=1; photon\_ratio=1;\\
do\_eloss=1; do\_espread=1; rndm\_seed=1; rndm\_load=0; rndm\_save=0;\}
}


\bibliographystyle{abbrv}
\bibliography{PairBackground_in_SiD.bib}

\begin{thebibliography}{10}

\bibitem{GuineaPigGrid}
{GuineaPig ASCII output files of the pair background for the ILC-500}.
\newblock /ilc/user/a/aschuetz/GuineaPig.

\bibitem{GuineaPigMan}
{GuineaPig Manual (SLAC NLC webpage version)}.
\newblock
  \url{http://www-sldnt.slac.stanford.edu/snowmass/Software/GuineaPig/gpman.pdf}.

\bibitem{geant_ref}
S.~Agostinell et~al.
\newblock {GEANT4: A Simulation toolkit}.
\newblock {\em Nucl. Instrum. Meth. A}, 506:250--303, 2003.

\bibitem{Aihara:2009ad}
H.~Aihara, P.~Burrows, M.~Oreglia, E.~L. Berger, V.~Guarino, J.~Repond,
  H.~Weerts, L.~Xia, J.~Zhang, Q.~Zhang, et~al.
\newblock {SiD Letter of Intent}.
\newblock 2009.

\bibitem{geant_ref2}
J.~Allison et~al.
\newblock {Geant4 developments and applications}.
\newblock {\em {IEEE Transactions on Nuclear Science}}, 53(1):270--278, Feb
  2006.

\bibitem{TDR}
{Behnke, T. et al. [ILC Collaboration]}.
\newblock {\em {The International Linear Collider Technical Design Report}}.
\newblock 2013.
\newblock arXiv:1306.6327 [physics.acc-ph].

\bibitem{ILC_TDR_4}
{Behnke, T. et al. [ILC Collaboration]}.
\newblock {\em {The International Linear Collider Technical Design Report --
  Volume 4, Detectors}}.
\newblock 2013.
\newblock arXiv:1306.6329 [physics.ins-det].

\bibitem{KPiX}
J.~Brau et~al.
\newblock {KPiX - A 1,024 channel readout ASIC for the ILC}.
\newblock {\em Proc. of the 2012 IEEE Symposium on Nucelar Science and Medical
  Imaging}, pages 1857 -- 1860, 2012.

\bibitem{Chen:1993dba}
P.~Chen, T.~L. Barklow, and M.~E. Peskin.
\newblock {Hadron production in gamma gamma collisions as a background for e+
  e- linear colliders}.
\newblock {\em Phys. Rev.}, D49:3209--3227, 1994.

\bibitem{Graf:2006ei}
N.~Graf and J.~McCormick.
\newblock {Simulator for the linear collider (SLIC): A tool for ILC detector
  simulations}.
\newblock {\em AIP Conf. Proc.}, 867:503--512, 2006.
\newblock [,503(2006)].

\bibitem{ref:Whizard}
W.~Kilian, T.~Ohl, and J.~Reuter.
\newblock Simulating multi-particle processes at lhc and ilc.
\newblock {\em Eur. Phys. J.}, C71:1742, 2011.

\bibitem{ref:antiDiD}
B.~Parker and A.~Seryi.
\newblock {Compensation of the effects of a detector solenoid on the vertical
  beam orbit in a linear collider}.
\newblock {\em Physical Review Special Topics -- Accelerators and Beams},
  8(4):041001, 2005.

\bibitem{Confluence}
A.~Schuetz and J.~Strube.
\newblock {Confluence page Simulation of the background events for the SiD
  detector'}.
\newblock
  \url{https://wikis.bris.ac.uk/display/sid/Simulation+of+the+background+events+for+the+SiD+detector}.

\bibitem{Schulte:1997nga}
D.~Schulte.
\newblock {\em {Study of Electromagnetic and Hadronic Background in the
  Interaction Region of the TESLA Collider}}.
\newblock PhD thesis, DESY, 1997.

\end{thebibliography}

\end{document}